# Phase-field modeling of $TiO_2$ nanocarving via reaction with hydrogen-bearing gas

Alireza Seifi, Sheikh Akbar, Yanzhou Ji*

*Department of Materials Science and Engineering, The Ohio State University, Columbus, OH 43210, USA*

**Abstract**

$TiO_2$ nanocrystals can be fabricated by carving the $TiO_2$ bulk polycrystals using reductive $H_2$-bearing gases, yielding single-crystal [001] nanowire arrays. However, the origin of the strongly anisotropic nanowire morphology during nanocarving remains largely unexplored. In this study, we formulate a 2-D phase-field model to investigate the $TiO_2$ morphology evolution during nanocarving processes. The model incorporates $TiO_2$ reduction reaction, $Ti^{3+}$ diffusion, and anisotropies in surface energy, diffusivity and reaction rate. Through systematic simulations in both single- and poly-crystals, we elucidate the roles of different anisotropy factors in nanocrystal morphologies at different carving stages, identifying the strong reaction rate anisotropy as the dominant factor for experimentally observed nanowire morphologies. We further explore the effect of grain misorientation angle, generating a nanocarving morphology map to guide grain orientation control. This study provides insights into microstructure evolution mechanisms during nanocarving and guidances for related microstructure control.

*Corresponding author. Email: ji.730@osu.edu

## Introduction

$TiO_2$ has been extensively used in industries due to its premium functionalities. For example, it can be used as chemical sensors for hydrocarbons, $CO_2$, $NO_x$, and $SO_x$ during petroleum production and refining. It can also be used as nanocatalysts for oxidative desulfurization (ODS) reactions to remove sulfur compounds from petroleum and its derivatives, preventing harmful $SO_x$ emissions. In practice, $TiO_2$ nanocrystals are usually synthesized to maximize their specific surface area, improve the contact between its surface atoms and the target molecules, thus enhancing its catalytic performance. The morphology of $TiO_2$ nanocrystals significantly affects its catalytic properties: the $TiO_2$-catalyzed ODS activity of diesel depends highly on the oxygen vacancy and $Ti^{3+}$ concentrations at the catalyst surface, which significantly varies with nanocrystal surface orientations. Therefore, it is critical to develop controllable synthesis strategies to fabricate $TiO_2$ nanocrystals with desirable morphologies and catalytic properties.

$TiO_2$ nanocrystals can be obtained using various synthesis strategies including sol-gel processes, precipitation and coprecipitation methods, hydrothermal methods, and so forth. However, these methods usually involve complicated interactions between source materials and physical/chemical fields, making it difficult to precisely control the $TiO_2$ nanocrystal morphology and the surface concentrations of catalytically active species.

Here we are particularly interested in the synthesis of $TiO_2$ nanocrystals using a surprisingly simple while effective "nanocarving" strategy, i.e., by using reductive $H_2$-bearing gases to carve the $TiO_2$ bulk crystals at elevated temperature (~700 °C). This approach has been experimentally demonstrated to yield single crystal $TiO_2$ nanowire arrays with [001] orientation with diameters of 20-50 nm and lengths of up to microns at the surface layer of the initial $TiO_2$ bulk polycrystal[1, 2], as shown in Figure 1(a). Moreover, this approach could become the building blocks for synthesizing more delicate nanocrystals with enhanced catalytic properties.

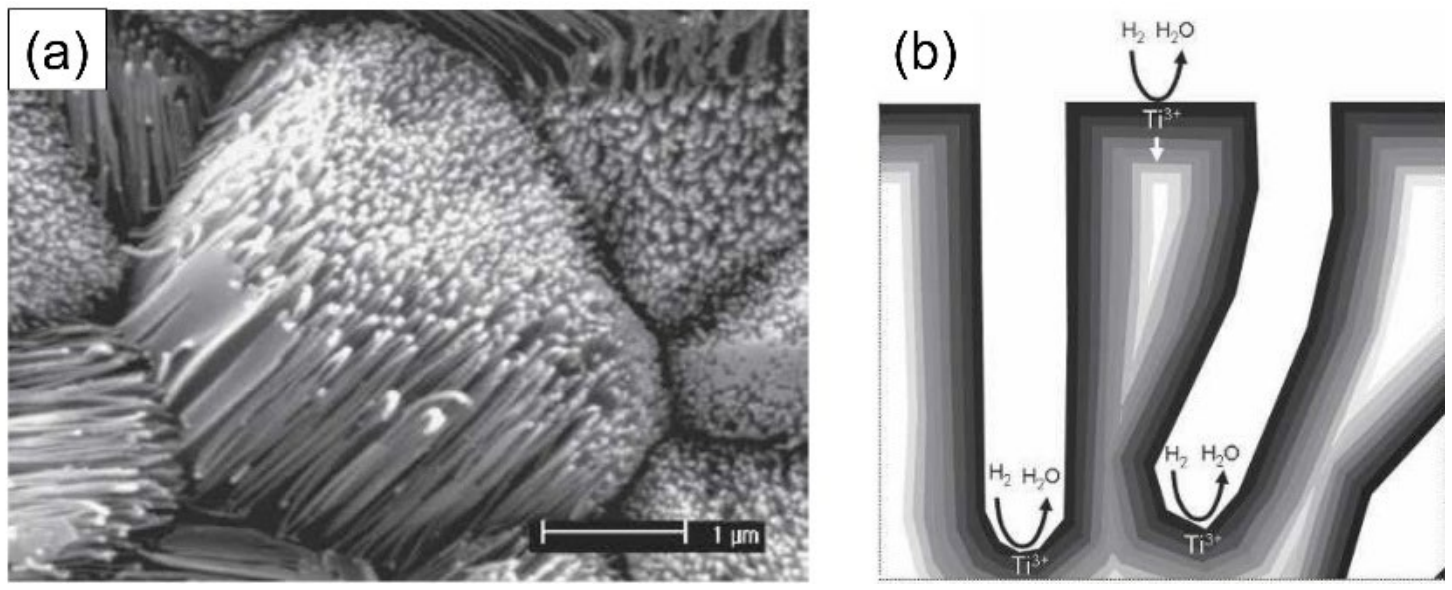

Figure 1: (a) Formation of $TiO_2$ nanofibers by nanocarving (adapted from [1]); (b) Illustration of the key reaction and diffusion processes during nanocarving (adapted from [1]).

The nanocarving of $TiO_2$ bulk crystals into $TiO_2$ nanowires is realized by two processes[1], (i) the reduction of $TiO_2$ by $H_2$ to form defective $TiO_{2-x}$ and (ii) the transport of $Ti^{3+}$ into the sinks deep inside $TiO_2$ bulk crystal via interstitial diffusion and grain-boundary (GB) diffusion. In the reduction reaction (i), one $H_2$ gas molecule combines with one oxygen anion $O^{2-}$ at the surface to form an $H_2O$ gas molecule, which will be removed from the $TiO_2$ surface; meanwhile, two electrons are transferred from the $H_2$ molecule to two $Ti^{4+}$ cations to form two $Ti^{3+}$ cations. The reaction can be expressed as

$$TiO_2 + xH_2 = TiO_{2-x} + xH_2O \qquad (1a)$$

Or

$$2Ti^{4+} + H_2 + O^{2-} = 2Ti^{3+} + H_2O \qquad (1b)$$

This reaction leads to redundant $Ti^{3+}$ cations at the crystal surface, which will be transported into the bulk crystal through process (ii) and eventually form the nanowire.

Although these two kinetic processes have been identified, their relative contributions to the overall nanocarving kinetics and the nanocrystal morphologies remain unclear. In particular, it is unclear why the nanowires are highly anisotropic and predominately [001]-oriented. To answer these questions, it is beneficial to dig deeper into the anisotropies involved in these processes. As shown in Figure 2(a), $TiO_2$ surface energies have been quantified as anisotropic [3]. The (001) surface energy is about twice of that of the (100) surface, leading to slightly anisotropic crystal shapes. Moreover, both the hydrogen reduction reaction rates and the $Ti^{3+}$ diffusion could be anisotropic. For example, the $Ti^{3+}$ bulk diffusivities in $TiO_{2-x}$ have been experimentally measured [4] to be higher in the (001) plane and lower along the [001] direction (Figure 2(b)). However, both these anisotropies would not lead to significant crystal shape anisotropies. Although strongly facet-dependent hydrogenation kinetics in anatase $TiO_2$ has been reported at 200 °C [5], such anisotropy has not been measured in rutile $TiO_2$ under the current carving conditions.

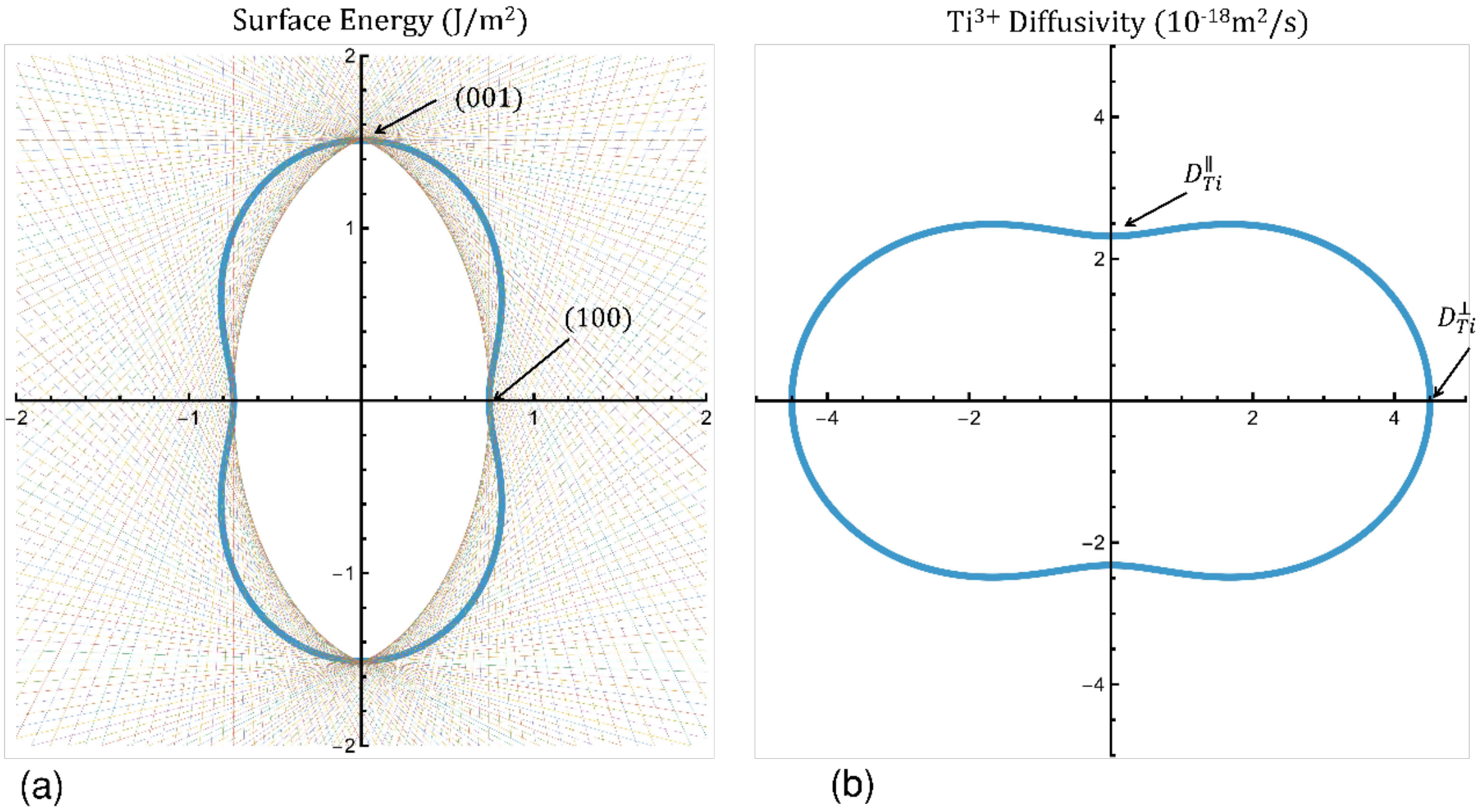

Figure 2: Anisotropies in (a) $TiO_2$ surface energy and (b) $Ti^{3+}$ diffusivity in the [100]-[001] plane. The dashed lines in (a) show the Wulff construction process. Their inner envelope shows the equilibrium crystal morphology.

On the other hand, theoretical and computational approaches can provide useful insights into the reaction and diffusion processes of $TiO_2$ nanocarving. The activation energies for $H_2$ dissociation and H transfer have been evaluated for different rutile $TiO_2$ surface orientations using density functional theory (DFT) calculations [6, 7]. Meanwhile, the Ti-O-H ReaxFF [8-10] has been developed and applied to study the reduction of $TiO_2$ nanosheets and nanoparticles by $H_2$, i.e., hydrogenation, with rutile, anatase and brookite crystal structures [11, 12]. These studies reveal the detailed molecular processes during $TiO_2$ reduction, the surface reactivity is shown to depend on crystal structures and surface orientations, and temperature, pressure and $H_2$ contents. However, the reaction rates have not been directly quantified in these studies.

Similar anisotropic material removal processes have also been observed in other crystalline oxides. During wet etching of ZnO, certain crystallographic facets form more readily, producing distinct pyramidal structures and revealing the orientation-dependent reactivity of the wurtzite crystal [13]. Likewise, single-crystal $\alpha$-$Al_2O_3$ (sapphire) exhibits strongly orientation-dependent wet-etching rates that determine the resulting etch-front and sidewall morphologies [14]. In $CeO_2$-based nanocrystals, facet-dependent wet etching kinetics was also observed, and the dominant facets for overall etching have been identified to be {100} for reduction and {111} for oxidation [15]. These observations indicate that crystallographically controlled surface reaction kinetics for material removal are general features of crystalline oxides rather than phenomena unique to $TiO_2$.

Overall, since nanocarving is a multiscale problem starting from bulk microscale crystals, and involves long-range diffusion, it is computationally impractical to simulate the entire carving process via atomistic simulations alone. Multiscale simulation approaches bridging physical and chemical phenomena at different length and time scales of the nanocarving process are indispensable. In particular, the phase-field (PF) approach, informed by atomistic and molecular simulations for the anisotropic energetics and kinetic rates, is promising in accurately predicting the nanocrystal morphology evolution. So far, the PF approach has not been directly applied to the nanocarving process, but it has been developed and applied to relevant problems including nanowire growth via the vapor-liquid-solid (VLS) approach [16], 2D materials growth via chemical vapor deposition (CVD) [17-19], and anisotropic pitting corrosion [20].

In this study, to provide insights into how the different anisotropy factors would interact to affect the $TiO_2$ nanocarving morphologies, we develop a 2-D PF model. The PF model involves all the key thermodynamic and kinetic processes during nanocarving, including bulk thermodynamic driving forces of the hydrogen reduction reaction, the defect thermodynamics of the rutile $TiO_2$ phase, along with the anisotropies in $TiO_2$ surface energies, hydrogen reduction reaction rates, and $Ti^{3+}$ diffusion. This PF model is a comprehensive extension of recent developments of PF modeling of stoichiometric reactions [21] and nonstoichiometric solution phase [22]. By varying the anisotropies, we generate a map of $TiO_2$ nanocarving morphologies, providing insights into which anisotropy factor dominates the morphology and leads to the experimentally observed nanowire morphologies. This study will trigger more systematic atomistic modeling and experimental characterization to validate kinetic anisotropies, and provide guidance for engineering the nanocrystal morphology via nanocarving and materials design.

## Results and Discussion

In this section, we implement our 2-D PF model in the xz plane and quantitatively investigate the effects of different anisotropies on nanocarving morphology. We start with simulations in single crystals to elucidate the roles of different anisotropy factors, and then extend our investigations to polycrystals. The simulation parameters are listed in Table 1. These parameters are normalized by a characteristic length $l_0 = 1\, nm$, a characteristic energy $E = 10^9\ J/m^3$, and a characteristic diffusivity $D = 4.50 \times 10^{-18}\ m^2/s$ in the simulations.

Table 1: PF simulation parameters

| Parameters | Explanations | Values [Units] | Source |
| --- | --- | --- | --- |
| $D^{*}_{ij,\mathrm{O}}$ (1000 K) | O self diffusivity | $\begin{pmatrix} 3\times10^{-18} & 0 \\ 0 & 2\times10^{-18} \end{pmatrix} \left[\frac{m^2}{s}\right]$ | [23, 24] |
| $D^{*}_{ij,\mathrm{Ti}}$ (1000 K) | Ti self diffusivity | $\begin{pmatrix} 4.50\times10^{-18} & 0 \\ 0 & 2.32\times10^{-18} \end{pmatrix} \left[\frac{m^2}{s}\right]$ | [4] |
| $V_m$ | Molar volume | $10^{-5}$ [$m^3$/mol] | |
| $w$ | Double-well height | $3.552\times10^{9}\ \left[\frac{J}{m^3}\right]$ | $w = 12\frac{\gamma}{\delta}$ |
| $\beta$ | Reaction rate anisotropy factor | 0, 10, $10^2$, $10^3$ | |
| $L_r$ | Reaction rate along [001] | $4.50416\ \times10^{-10}\ \left[\frac{m^3}{J.s}\right]$ | Estimated |
| $\delta$ | Interface thickness | $\begin{pmatrix} 2.5\times10^{-9} & 0 \\ 0 & 5.1013\times10^{-9} \end{pmatrix} [m]$ | Estimated |
| $\kappa_{ij}$ | Gradient energy coefficient | $\begin{pmatrix} 2.775\times10^{-9} & 0 \\ 0 & 1.1554\ \times10^{-8} \end{pmatrix} \left[\frac{J}{m}\right]$ | $\kappa = \frac{3}{2}\gamma\delta$ |
| $\gamma_{ij}$ | Interfacial energy | $\begin{pmatrix} 0.74 & 0 \\ 0 & 1.51 \end{pmatrix} \left[\frac{J}{m^2}\right]$ | [3] |

***Nanocarving in single crystals: elucidating the anisotropy effects***

The morphology of nanocarved grains is dependent on several competing anisotropic factors: interfacial energy, reaction rate, and diffusivity anisotropy. To fully understand their roles in nanocarving morphology, we start with 2-D phase-field simulations in single crystals with only one initial pit and different combinations of the anisotropy factors. As shown in Figure 3(a), the computational domain consists of $200\ \times 160$ grid points with the spatial discretization $dx = dy = 1.0\ nm$ and a time step $dt = 4.5\times10^{-4}s$. The initial pit is a semicircle with a diameter of 5 nm at the top center of the simulation domain. The simulations are performed under $T$=1000 K, $p_{H_2O}/p_{H_2} = 5\ \times10^{-9}$, and an initial Ti composition of $x^0_{Ti} = 0.333334$ in the $TiO_2$ matrix. This initial composition and $p_{H_2O}/p_{H_2}$ ratio ensure the reaction has sufficient driving force to take place (See

Figure S2 in Supplementary Materials A.4), while not deviating too much from the ideal $TiO_2$ stoichiometry. Zero-flux (Neumann) boundary conditions are applied on all the external boundaries for both order parameters and $x_{Ti}$, preventing mass exchange across those boundaries. This will lead to ~3.1 vol.% $TiO_2$ carved away at equilibrium.

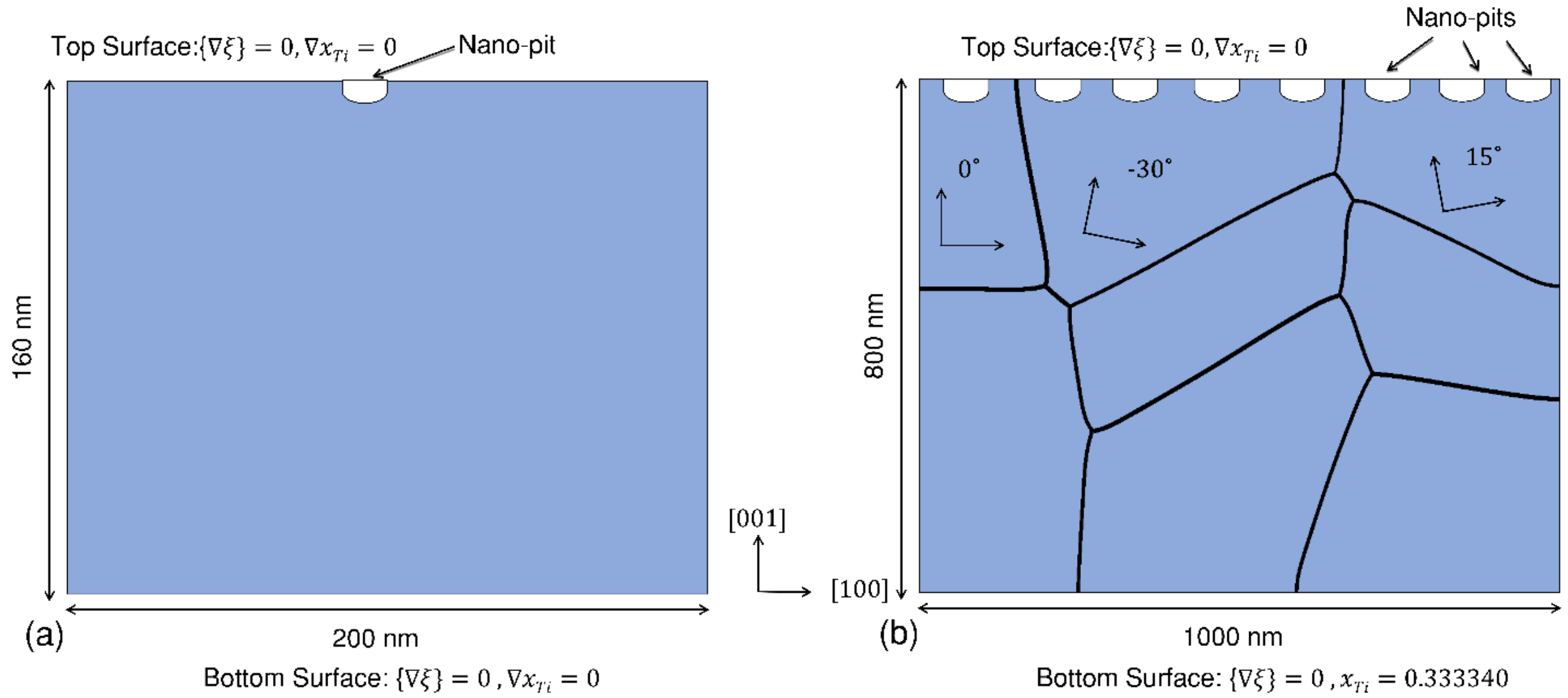

Figure 3: Initial and boundary conditions for 2-D PF simulations in the [100]-[001] plane: (a) Single crystal; (b) Polycrystal. The axes and angles inside each surface grain in (b) indicate the grain misorientation with respect to the global [100]-[001] reference frame.

The simulated carving morphologies under different combinations of the anisotropy factors are summarized in Figure 4. The carving morphology evolution is further quantified by the aspect ratio of $\eta = \frac{l_y}{l_x/2}$, where $l_x$ and $l_y$ are the diameter and depth of the pit, respectively, as shown in Figure 4(a).

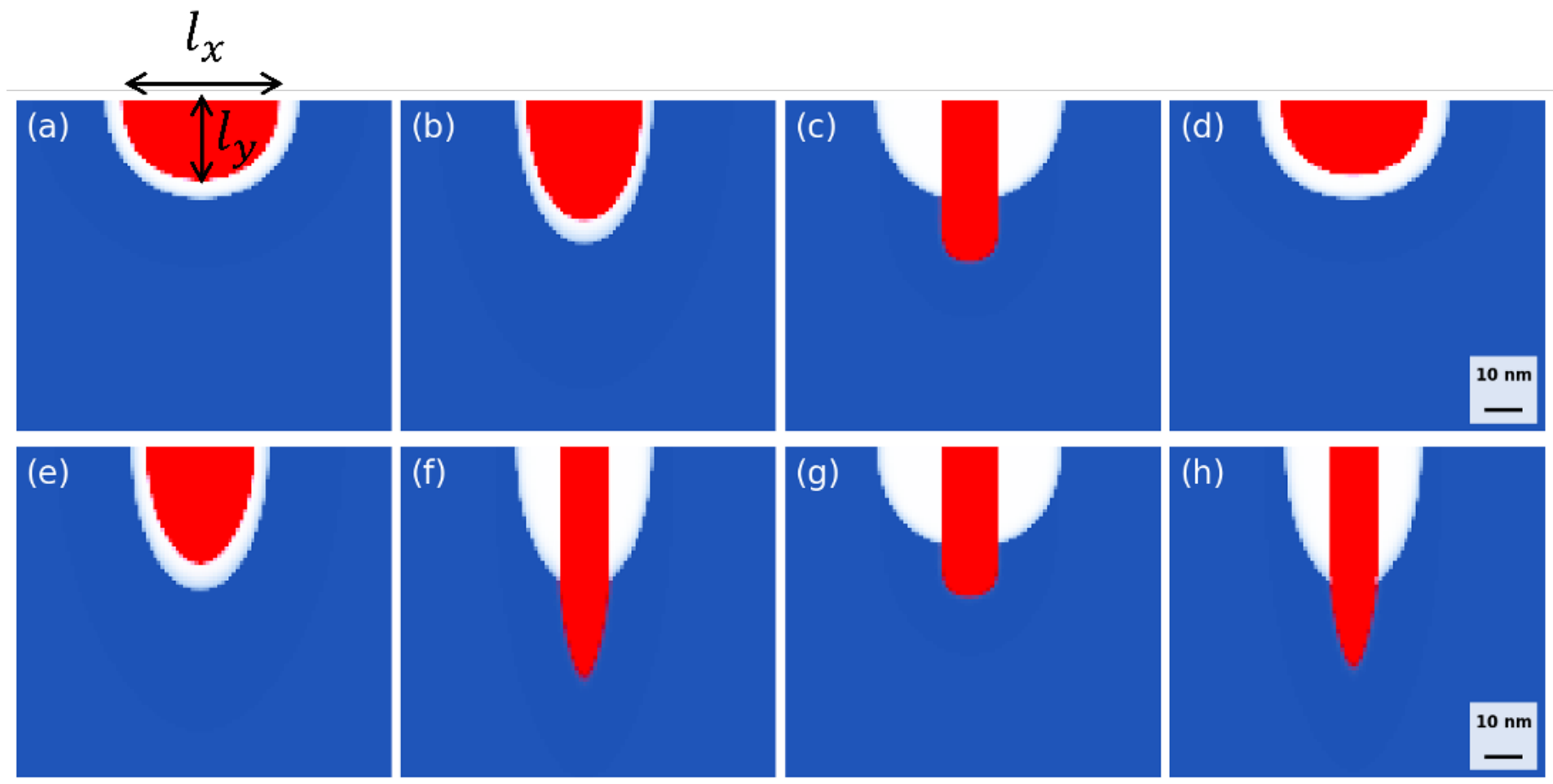

Figure 4: Carving morphology of the single crystal at 0.30 minutes (red pit) and 74 minutes (white pit) for different anisotropy combinations: (a) fully isotropic, (b) only interfacial energy (γ) anisotropy, (c) only reaction rate ($L_r$) anisotropy, (d) only diffusivity (D) anisotropy, (e) combined interfacial energy and diffusivity anisotropies, (f) combined interfacial energy and reaction rate anisotropies, (g) combined diffusivity and reaction rate anisotropies, (h) fully anisotropic.

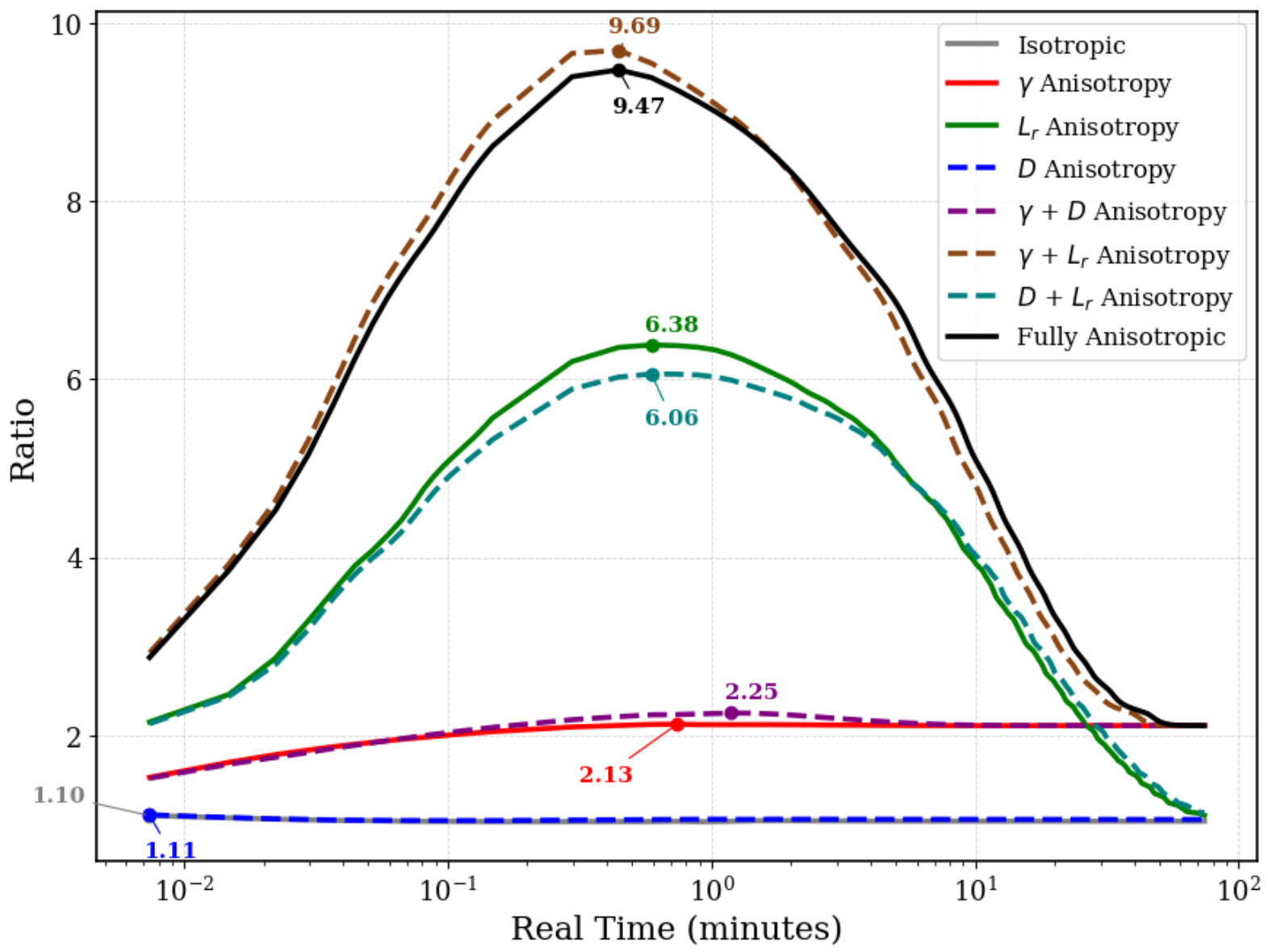

Figure 5: Aspect ratio (η) variation vs. the logarithm of real time for different isotropic and anisotropic cases. The peak ratio for each case is also shown.

We start our analysis from the fully isotropic case. As shown in Figure 4(a), the carving is isotropic, and the pit shape remains a semi-circle. Meanwhile, the depth vs. radius aspect ratio of the pit stays almost 1 throughout the carving process (Figure 5: "Isotropic").

We then study the effects of individual anisotropy factors on carving morphology. As shown in Figure 4(b), the interfacial energy anisotropy causes more elongations along the y-axis, eventually leading to an aspect ratio of ~2 (Figure 5: γ Anisotropy), which is well consistent with the prediction from the Wulff construction, i.e., $\frac{\gamma_{yy}}{\gamma_{xx}} = \frac{1.51}{0.74} = 2.04$. Reaction rate anisotropy alone with $\beta$=10 in Figure 4(c) shows higher initial penetration due to higher $L_r$ along the y-direction. According to Figure 5: $L_r$ Anisotropy, the aspect ratio first increases to ~6.5 and then gradually decreases to 1, eventually becoming almost isotropic. This is because its initial increase is dominated by the anisotropic reaction kinetics, but if given sufficiently long time, thermodynamics would dominate, and the pit becomes less elongated. There is a transition from a kinetics-dominated regime to a thermodynamics-limited regime. On the other hand, diffusivity anisotropy (Figure 4(d), Figure 5: D Anisotropy) has negligible effects, and the morphology tends to stay semi-circle, similar to the fully isotropic case, even though $D_{xx}^* > D_{yy}^*$ for both Ti and O.

Next, we combine each of these factors in pairs (Figure 4(e-g)). By comparing Figure 4(e) with Figure 4(b) and Figure 4(g) with Figure 4(c), we see that the effect of diffusivity anisotropy is limited. From a more quantitative analysis in Figure 5, we find that the "D+Lr Anisotropy" curve has a slightly lower peak than the "Lr Anisotropy" curve, which can be attributed to the competition between the reaction rate and diffusivity anisotropies: the reaction rate anisotropy tends to deepen the pit, while the diffusivity anisotropy tends to widen the pit. On the other hand, by comparing Figure 4(f) with Figure 4(b) and Figure 4(c), and by the quantitative analysis in Figure 5, we find that the "γ+Lr Anisotropy" case gives the highest peak $\eta$ ratio of ~9.6, much higher than the individual effects of interfacial energy anisotropy (peak $\eta$~2) and reaction rate anisotropy (peak $\eta$~6.5). This can be attributed to their synergistic effect: both of them tend to deepen the pit. Therefore, interfacial energy and reaction rate anisotropies synergize, while both compete with the diffusivity anisotropy to determine the early and intermediate carving morphology; during late carving stages, however, both reaction rate and diffusivity anisotropy effects gradually vanish due to their kinetic nature, while the interfacial energy anisotropy persists to determine the final equilibrium carving morphology due to its thermodynamic nature. Ultimately, with all anisotropies considered, the carving morphology evolution is close to the "γ+Lr Anisotropy" case (Figure 4(h) vs. (f)), with a slightly lower peak $\eta$ ratio of ~9 due to the inclusion of diffusivity anisotropy (Figure 5: Fully Anisotropic).

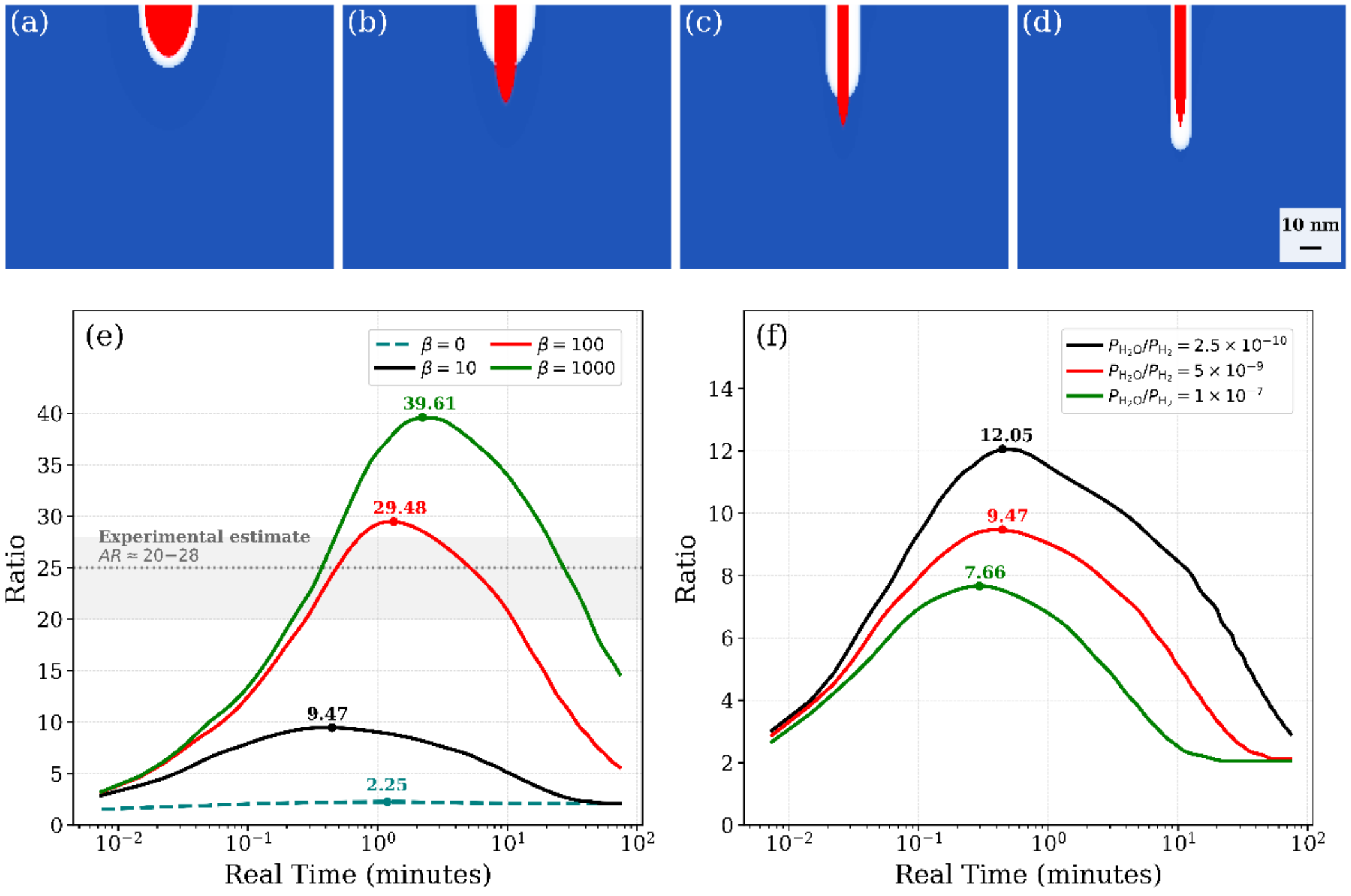

Figure 6: Carving morphology of the single crystal at 0.30 minutes (red pit) and 74 minutes (white pit) for different $\beta$ values: (a) $\beta$=0 (no reaction rate anisotropy), (b) $\beta$=10, (c) $\beta$=100, (d) $\beta$=1000. Aspect ratio (η) variation vs. the logarithm of real time for different $\beta$ values (e) and different $P_{H2O}$ / $P_{H2}$ values (f).

Based on these simulation results, we find that the reaction rate anisotropy critically governs the carving morphology at early and intermediate carving stages. Unfortunately, unlike interfacial energy and diffusivity anisotropies, the detailed reaction rates at different rutile $TiO_2$ surfaces have not been reported in the literature. To further quantify its anisotropy effect, we perform a systematic parametric study by varying the $\beta$ values in phase-field simulations while fixing other anisotropies. Generally, as shown in Figure 6(a-d), as $\beta$ increases, the penetration becomes more rapid and deeper in the y-direction, while the widening of the pit becomes slower. As further quantified in Figure 6(e), all the aspect ratio curves except $\beta$=0 (no reaction rate anisotropy) show peaks in the middle, indicating the dominance of kinetics at early and intermediate carving stages. As $\beta$ increases, the peak ratio increases from ~10 ($\beta$=10) to ~40 ($\beta$=1000). At late carving stages, the aspect ratio would all decay and eventually converge to ~2 due to the dominance of thermodynamics; but the time needed for this convergence increases with the increase of $\beta$. In comparison, the average aspect ratio of the nanofibers is approximately 25 based on estimations from the experimental images in Figures 2E and 4A in [2], as shown by the black dotted line in Figure 6(e). The variations are largely due to nonuniform diameters of the nanofibers.

The physical origin of reaction-rate anisotropy can be related to the different atomic structures and coordination environments of the exposed rutile surfaces, which lead to orientation-dependent interactions with hydrogen. $TiO_2$ hydrogenation contains multiple steps involving $H_2$ adsorption and dissociation, Ti–O bond breaking, $H_2O$ formation and desorption, and subsequent defect redistribution and structural reconstruction [12]; each of these steps can be anisotropic. For example, $H_2$ dissociation barriers have been evaluated to be 0.50, 0.56, 0.79, and 1.08 eV on rutile (110), (001), (101), and (100), respectively [25], demonstrating a strong crystallographic dependence. Moreover, orientation-dependent hydrogen adsorption and preferential hydrogenation-induced disorder along the facets of rutile nanoparticles have been captured with ReaxFF molecular dynamics (MD) simulations [12]. Experimentally, strongly facet-dependent hydrogenation kinetics in anatase $TiO_2$ has been reported at 200 °C, with the rates following the order (100) > (101) > (001) due to anisotropic oxygen-vacancy diffusion [5]. Although quantitative facet-specific overall reaction rates have not yet been reported under rutile nanocarving conditions, these atomistic and experimental studies provide a physical basis for crystallographically dependent $TiO_2$ reduction kinetics. The present simulations further show that, when reaction-rate, diffusion, and interfacial-energy anisotropies are all considered, reaction-rate anisotropy has the strongest influence on nanofiber formation and aspect ratio. This result therefore provides a mechanistic prediction that facet-dependent reduction kinetics are a primary factor governing morphology selection during rutile $TiO_2$ nanocarving.

The processing atmosphere also influences the nanocarving morphology through the thermodynamic driving force for $TiO_2$ reduction. In the present model, the $H_2$ and $H_2O$ partial pressures enter the governing equation through the ratio $p_{H_2O}/p_{H_2}$. Decreasing this ratio produces a more reducing atmosphere and increases the reaction driving force, whereas increasing the ratio decreases the driving force and suppresses further reduction, as shown in Figure S2. To examine the corresponding morphological effect, additional closed single-crystal simulations were performed at $p_{H_2O}/p_{H_2} = 2.5 \times 10^{-10}$ and $10^{-7}$, while keeping the remaining model parameters unchanged. As shown in Figure 6(f), decreasing $p_{H_2O}/p_{H_2}$ produces a larger aspect ratio at the same processing time. Thus, under otherwise identical conditions, a stronger reducing atmosphere promotes a more elongated transient morphology by increasing the chemical driving force and enhancing the influence of the anisotropic reaction kinetics on interface evolution. At longer times, the differences among the aspect-ratio curves progressively decrease, indicating a transition from kinetically dominated morphology evolution toward increasing thermodynamic control.

Temperature provides an additional processing variable that affects both reaction and diffusion kinetics. Yoo et al. experimentally observed Arrhenius behavior of the rutile nanocarving rate between 682 and 772 °C, with an activation energy of 57.3 ± 9.0 kJ/mol [1]. Reactive MD simulations have also shown that increasing temperature and $H_2$ content promotes $TiO_2$ reduction. On the other hand, sufficiently high external pressure can slow down atomic diffusion and structural reconstruction [12]. These results indicate that processing parameters such as temperature and external pressure can influence both the overall nanocarving rate and the resulting nanofiber morphology through their effects on the relative contributions of kinetic and thermodynamic processes.

From an experimental perspective, reaction-rate anisotropy can be realized by controlling the crystallographic orientation and facet exposure of rutile $TiO_2$ prior to nanocarving treatment. Oriented rutile single crystals, textured polycrystalline materials, or facet-engineered rutile structures can be used to expose selected crystallographic surfaces to the hydrogen-containing atmosphere. Preferential exposure of specific rutile facets has been demonstrated experimentally [26, 27], providing a practical route for controlling the initial surface orientations involved in nanocarving. Because different rutile facets exhibit different activation barriers of hydrogenation [25], the relative difference between their reaction rates is also expected to vary with temperature according to Arrhenius kinetics. Hydrogen treatment and $H_2$ pressure can further modify hydrogen-related defects and oxygen-vacancy populations [28], which may influence the reduction kinetics. In addition, the $p_{H_2O}/p_{H_2}$ ratio controls the thermodynamic driving force for $TiO_2$ reduction and therefore the overall nanocarving rate [1]. Therefore, experiments combining controlled rutile orientations with systematic variations of temperature and gas pressures provide a practical route for tuning and directly testing the reaction-rate anisotropy predicted by the present model.

***Nanocarving in polycrystals***

With this new understanding of the anisotropy effects, we further extend our simulations to polycrystals with multiple initial pits. The initial $TiO_2$ polycrystalline structure is obtained using an in-house PF grain growth simulation [29] with 8 different grain orientations. This polycrystal microstructure is assumed static during carving. We further assume that carving takes place only in grains connected to the top surface and stops when touching GBs. We also apply a Dirichlet boundary condition for $x_{Ti}$ at the bottom surface with $x_{Ti} = 0.33334$ to effectively consider a thick sample. The initial pits are assigned with equal size and spacing at the top surface, as shown in Figure 3(b). The simulation domain is discretized into $1000 \times 800$ grid points. Other simulation set-ups are the same as the single-crystal simulations discussed above.

We first perform phase-field simulations with all anisotropies considered and vary the reaction rate anisotropy factor $\beta$, as shown in Figure 7. At $\beta$=0, the carving morphology is dominated by interfacial energy anisotropy with low aspect ratios. As a result, the initial pits are easily connected with each other, and we could not see experimentally observed fine arrays of nanofibers with high aspect ratios and uniform length and diameter. As $\beta$ increases, the carved nanowires become thinner with less variations in length and diameter. In particular, the thickening of the initial pits is greatly limited, suppressing their connection. Ultimately, when $\beta$=100 and 1000, we get thinner nanowire arrays with uniform length and diameter that can sustain throughout intermediate carving stages, consistent with observations. This trend indicates that anisotropy of the reaction rate is the dominant factor leading to the experimentally observed nanowire morphologies. Its mechanism is primarily to suppress the thickening of initial pits to avoid their connection.

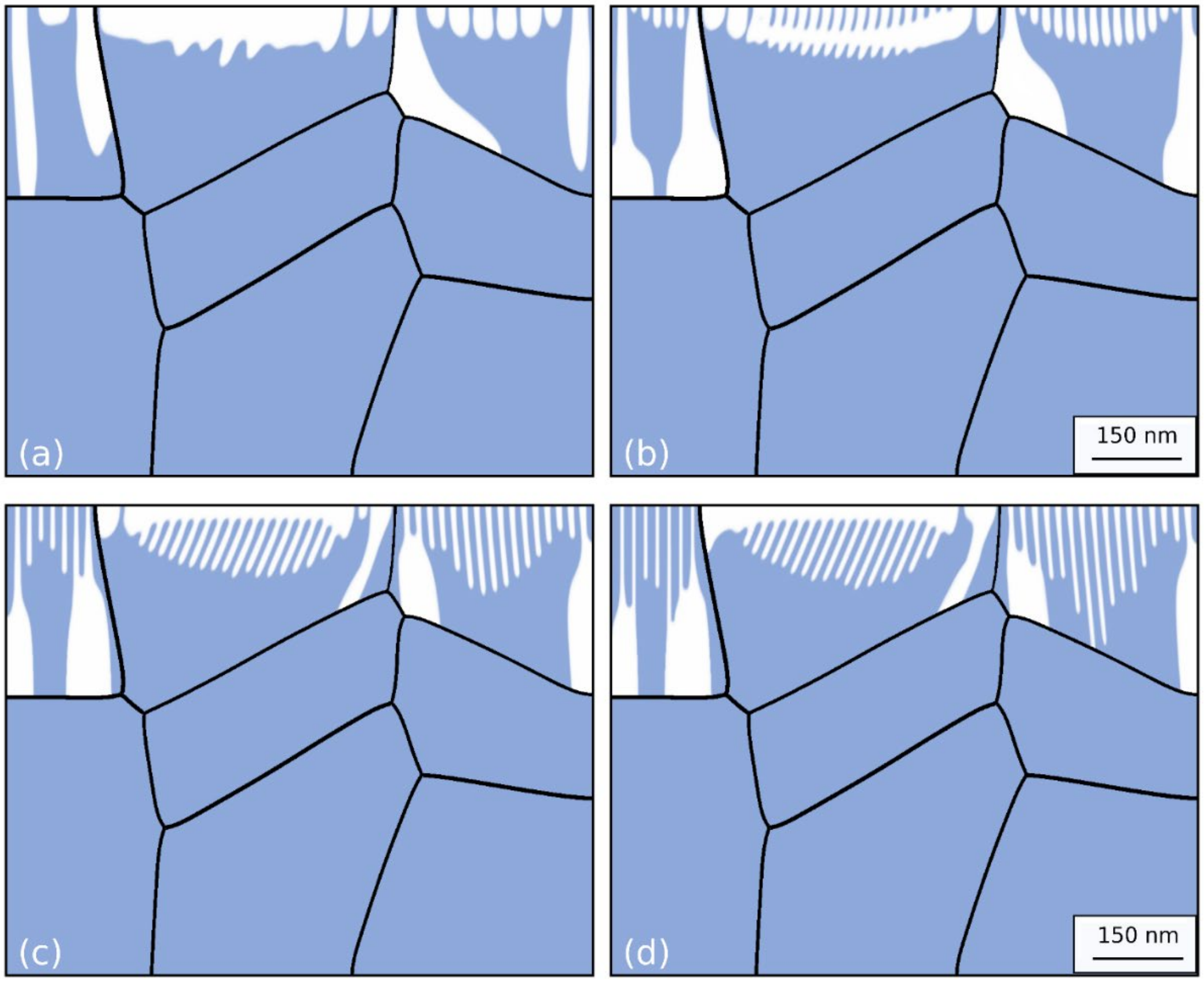

Figure 7: Carving morphology of the polycrystal at 0.3 min for different $\beta$ values: (a) $\beta$=0 (no reaction rate anisotropy), (b) $\beta$=10, (c) $\beta$=100, (d) $\beta$=1000. White color shows carved regions of $TiO_2$ grains surfaces (gas phase) and blue color shows $TiO_2$ phase.

From Figure 7(c-d), we also observe that the nanowire orientations are different in different grains, and different grains may experience different extents of carving. To further explore this effect, we only consider carving in one given grain – the grain at the top center of the simulation domain – and vary its grain orientation. We fix $\beta$=1000 and vary the grain orientation angle from -90° to 90° at an interval of 5°. As summarized in Figure 8, the resulting carving morphologies can be classified into different regions depending on the grain misorientation angle $\theta^0$, which measures the angle of the [001] crystal axis of the grain with respect to the y-direction of the lab reference frame. In region A, where $\theta^0$ is between -5° and 5°, only a small portion of the surface is completely carved, and the sharpest and the largest number of nanowires exist. As $|\theta^0|$ increases and Region B (5°~25°) and -B (-35°~-5°) emerge, we can observe that the number of thin nanowires decreases and a larger volume fraction of the grain is carved. Meanwhile, these thin nanowires are on top of a big "trunk" at the bottom of the grain. In Region C (25°~50°) and -C (-60°~-35°), we only see one nanowire, which is the "trunk" in regions B and -B, and most of the grain is carved. Finally, in Region D ($\theta^0 > 50°$) and -D ($\theta^0 < -60°$), there are no nanowires, and the surface of the grain is uniform carved with a flat surface. Of course, the specific boundaries of these different regions depend on the grain location and the GB orientations. This trend shows that the nanocarving morphology is highly dependent on grain orientation. By adjusting the grain orientation angle, we can obtain the desired morphology of the carved areas. In particular, if we desire higher nanowire density and uniform nanowire aspect ratio, it would be beneficial to achieve $TiO_2$ polycrystals with strong <001> texture.

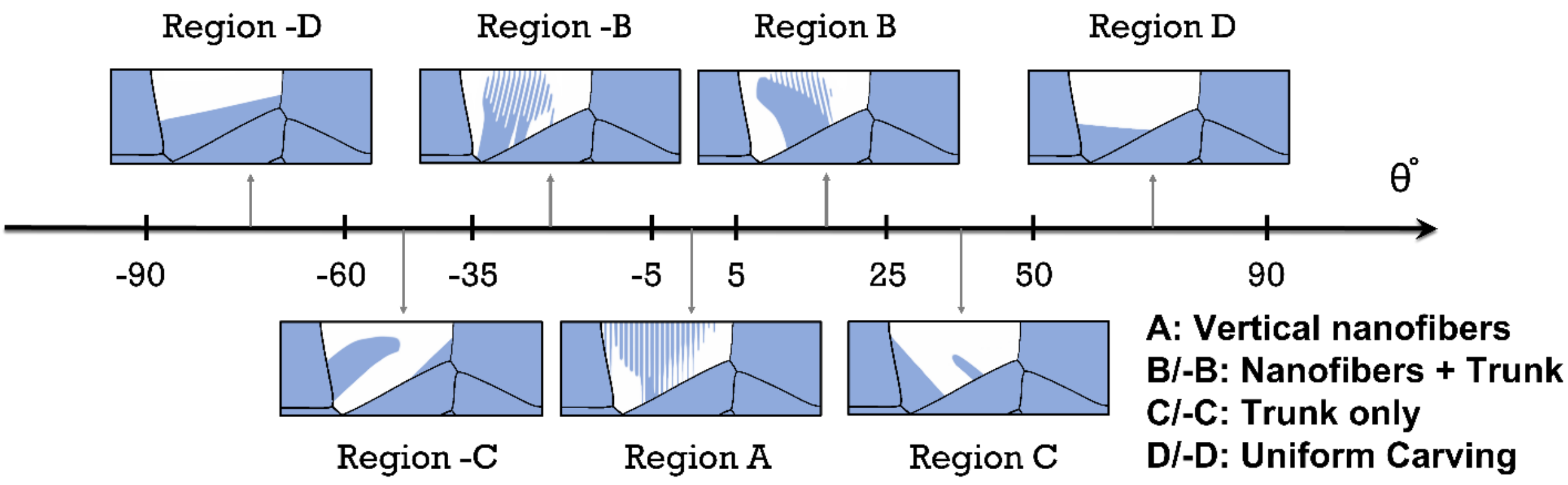

Figure 8: Carving morphology map of a specific grain in the polycrystal systems under varying grain misorientation angles at 0.3 min.

It should be noted that grain boundaries may further influence the overall nanocarving kinetics through enhanced solid-state transport. Yoo et al. showed that the nanocarving rate decreases with increasing rutile grain size and concluded that grain-boundary diffusion is an important transport mechanism [1], which we will investigate in detail in the future. In the present model, the carving front is additionally assumed to remain within the parent grain. This assumption is motivated by experimental observations of crystallographically distinct nanofiber arrays within individual rutile grains [1, 2], but the experiments do not establish grain boundaries as absolute barriers to carving. Therefore, possible grain-boundary reactions and cross-boundary propagation are not explicitly considered and represent an idealization in the present model.

## Conclusions

In this study, we investigated the $TiO_2$ nanowire formation and evolution during the carving process using a 2-D PF model of stoichiometric reaction and nonstoichiometric solution phases. Our 2-D PF model consists of all the key thermodynamic and kinetic processes during nanocarving, as well as their anisotropies, to capture the evolution of $TiO_2$ nanocrystal morphology. The results demonstrate that:

- The reaction rate anisotropy dominates the nanocarving aspect ratio at early and intermediate stages. Strong reaction rate anisotropy can lead to experimentally observed highly anisotropic [001] nanowire array morphology. Specifically, reaction rate anisotropy synergizes with surface energy anisotropy to promote the nanocarving aspect ratio along [001], while it competes with diffusivity anisotropy, which promotes nanocarving perpendicular to [001]. At late carving stages, the effect of reaction rate anisotropy decays due to its kinetic nature, while the surface energy anisotropy dominates the nanocarving morphology due to its thermodynamic nature.
- When multiple initial pits exist, the high reaction rate anisotropy would effectively promote deepening of the pits while suppressing their thickening, limiting their connection, and thus maintaining the nanowire carving morphologies with uniform diameter distributions during initial and intermediate carving stages.

- Nanocarving morphologies are also affected by grain orientation. According to our nanocarving morphology map as a function of grain misorientation angle, as $|\theta^0|$ increases, the number of thin nanowires decreases, and the grain is carved in a larger volume fraction.

Overall, we elucidated the role of different anisotropy factors in nanocarving morphology evolution. Our findings also suggest that we can adjust nanocarving morphology by controlling the different anisotropies and grain orientations to acquire desired nanocrystal morphologies for different applications. Our model can also be extended to the carving processes of other materials. Nevertheless, quantification of reaction rate anisotropies is recommended via ReaxFF MD simulation to further validate this anisotropy effect.

**Methods**

Here we develop a 2-D PF model including all the key thermodynamic and kinetic processes during nanocarving to predict the $TiO_2$ nanocrystal morphology evolution. The PF model combines the recent PF model of stoichiometric compounds and solution phases [21], and the PF model of nonstoichiometric solutions [22], where we treat the rutile $TiO_2$ as a nonstoichiometric solution phase and the gas phase as a stoichiometric phase. The $TiO_2$ microstructure will be described by three sets of PF variables: (1) the grain order parameters $\{\phi_i\}$ for initial $TiO_2$ grains with different orientations, which take $\phi_i = 1$ and $\phi_{j\neq i} = 0$ in the $i$th grain; (2) the local extent $\{\xi_i\}$ of the $TiO_2$ reduction reaction inside the $i$th grain, which also distinguishes the rutile $TiO_2$ ($\xi_i$=*0*) and gas ($\xi_i$=*1*) phases; (3) the local Ti composition in the rutile phase in atom ratio, $x_{Ti}$, which accounts for both $Ti^{3+}$ and $Ti^{4+}$ species Since the nanocarving temperature is much lower than the $TiO_2$ melting point, we will neglect the $TiO_2$ grain growth during nanocarving and assume $\{\phi_i\}$ are static. We also assume that nanocarving initiates from the $TiO_2$ surface and stops at GBs, i.e., there is only one active $\{\xi_i\}$ inside each grain and no interactions are considered among different $\{\xi_i\}$.

The total free energy $F$ of the system is formulated as

$$F = \int_V (f_{bulk} + f_{int}) dV \tag{2}$$

Where $f_{bulk}$ and $f_{int}$ are the bulk and interfacial energy densities, respectively. $f_{bulk}$ is formulated as

$$f_{bulk} = \frac{1}{V_m}\left\{\left[1 - \sum_i h(\xi_i)\right] G_m^{rutile}(x_{Ti}, T) + \sum_i h(\xi_i)\, G_m^{gas}\left(p_{H_2}, p_{H_2O}, T\right)\right\} \tag{3}$$

Where $V_m$ is the molar volume, $h(\xi) = 6\xi^5 - 15\xi^4 + 10\xi^3$ is an interpolation, $p_{H_2}$ and $p_{H_2O}$ are the partial pressures of $H_2$ and $H_2O$ in the gas phase, respectively, $T$ is the temperature, $G_m^{rutile}$ and $G_m^{gas}$ are the molar Gibbs free energies of the rutile and gas phases, respectively, taken from existing thermodynamic databases [30, 31]. The detailed thermodynamic formulations, including the molar Gibbs free energies, chemical potentials and reaction driving forces, are included in Supplementary Materials A; the corresponding thermodynamic functions are shown in Figure S1.

In particular, the rutile $TiO_2$ phase is described by the sublattice model $(Ti^{3+}, Ti^{4+})_1(O^{2-}, Va)_2$ to account for its nonstoichiometry due to the formation of $Ti^{3+}$ and oxygen vacancy (Va) during $TiO_2$ reduction reactions [20, 30]. Its Gibbs free energy per mole of formula unit is formulated as a function of sublattice site fractions $y^{I}_{Ti^{3+}}$, $y^{I}_{Ti^{4+}}$, $y^{II}_{O^{2-}}$ and $y^{II}_{Va}$ (see Supplementary Materials A.1) which can be converted to the Gibbs free energy per mole of atoms as a function of Ti composition, according to our existing approach [22, 32]. This conversion process is detailed in Supplementary Materials A.3, and the resulting Gibbs free energy and sublattice site fractions are shown in Figure S1 of the Supplementary Materials. In particular, $x_{Ti}$ is related to these site fractions via $x_{\mathrm{Ti}} = \frac{y^{I}_{\mathrm{Ti}^{3+}}+y^{I}_{\mathrm{Ti}^{4+}}}{y^{I}_{\mathrm{Ti}^{3+}}+y^{I}_{\mathrm{Ti}^{4+}}+2\,y^{II}_{\mathrm{O}^{2-}}} = \frac{1}{1+2\,y^{II}_{\mathrm{O}^{2-}}}$. $f_{int}$ is formulated to account for the anisotropic $TiO_2$ surface energies,

$$f_{int} = w\sum_{p}\xi_p^2\left(1-\xi_p\right)^2 + \frac{1}{2}\sum_{p}\kappa_{ij}\left(\nabla_i\xi_p\right)\left(\nabla_j\xi_p\right) \tag{4}$$

Where $w$ is the double-well height, and $\kappa_{ij}$ is the anisotropic gradient energy coefficient matrix, which depends on the anisotropic $TiO_2$ surface energy. $w$ and $\kappa_{ij}$ can be analytically obtained from the anisotropic interfacial energies and thicknesses [21].

The governing evolution equations of the PF variables are [21]

$$\frac{\partial\xi_p}{\partial t} = -\frac{V_m}{RT}k\left(\theta_p-\theta_p^0\right)\left[\frac{1}{V_m}\frac{\partial h}{\partial\xi_p}\Delta G_m^r + \frac{\delta\int_V f_{int}dV}{\delta\xi_p}\right] \tag{5}$$

$$\left[1-\sum_{i}(\xi_i)\right]\frac{\partial x_{Ti}}{\partial t} = \nabla_i\left[M_{ij}(x_{Ti})\nabla_j\left(\frac{\partial G_m}{\partial x_{Ti}}\right)\right] + x_{Ti}\sum_{i}\frac{\partial h}{\partial\xi_i}\frac{\partial\xi_i}{\partial t} \tag{6}$$

where Eq. (5) accounts for the reduction reaction, in which $\Delta G_m^r = 2\mu_{Ti^{3+}}^{rutile}(x_{Ti},T) + \mu_{H_2O}^{gas}\left(p_{H_2O},T\right) - 2\mu_{Ti^{4+}}^{rutile}(x_{Ti},T) - \mu_{H_2}^{gas}\left(p_{H_2},T\right) - \mu_{O^{2-}}^{rutile}(x_{Ti},T)$ is the Gibbs energy change of the reduction reaction, with $\mu_i^j$ being the chemical potential of species *i* in phase *j*, derived from the $G_m^{rutile}$ and $G_m^{gas}$ expressions in the thermodynamic database, as detailed in Supplementary Materials A.4 and A.1-A.2; $\frac{\delta\int_V f_{int}dV}{\delta\xi_p} = 2w\xi_p\left(1-\xi_p\right)\left(1-2\xi_p\right) - \kappa_{ij}\nabla_i\nabla_j\xi_p$ is the variational derivative of the total interfacial energy; $k\left(\theta_p-\theta_p^0\right)$ is the orientation-dependent reaction rate, which is given by the following equation [33]:

$$k(\theta_p-\theta_p^0)=\frac{RT}{V_m}\frac{L_r}{1+\beta}\begin{cases}1+\beta\cos(\theta_p-\theta_p^0), & -\frac{\pi}{2}+\phi_0\le\theta_p-\theta_p^0\le\frac{\pi}{2}-\phi_0,\\ 1+\frac{\beta}{\sin\phi_0}+\frac{\beta\cos\phi_0}{\sin\phi_0}\sin(\theta_p-\theta_p^0), & -\frac{\pi}{2}\le\theta_p-\theta_p^0<-\frac{\pi}{2}+\phi_0,\\ 1+\frac{\beta}{\sin\phi_0}-\frac{\beta\cos\phi_0}{\sin\phi_0}\sin(\theta_p-\theta_p^0), & \frac{\pi}{2}-\phi_0\le\theta_p-\theta_p^0\le\frac{\pi}{2},\\ 1-\beta\cos(\theta_p-\theta_p^0), & -\pi\le\theta_p-\theta_p^0<-\frac{\pi}{2}-\phi_0,\\ 1+\frac{\beta}{\sin\phi_0}+\frac{\beta\cos\phi_0}{\sin\phi_0}\sin(\theta_p-\theta_p^0), & -\frac{\pi}{2}-\phi_0\le\theta_p-\theta_p^0<-\frac{\pi}{2},\\ 1-\beta\cos(\theta_p-\theta_p^0), & \frac{\pi}{2}+\phi_0<\theta_p-\theta_p^0\le\pi,\\ 1+\frac{\beta}{\sin\phi_0}-\frac{\beta\cos\phi_0}{\sin\phi_0}\sin(\theta_p-\theta_p^0), & \frac{\pi}{2}<\theta_p-\theta_p^0\le\frac{\pi}{2}+\phi_0.\end{cases} \quad (7)$$

With $\theta_p=\arctan(\nabla_y\xi_p/\nabla_x\xi_p)$ being the surface normal orientation angle, $\theta_p^0$ being the grain orientation angle with respect to the lab reference frame, and $\phi_0=10^{-5}\pi$ being a small regularization angle. $L_r$ is an interfacial kinetic coefficient related to the reaction rate, and $\beta$ is a coefficient controlling the reaction rate anisotropy, so that the reaction rate is $\frac{RTL_r}{V_m}$ along [001] direction ($\theta_p-\theta_p^0=0$) and about $\frac{RT}{V_m}\frac{L_r}{1+\beta}$ along [100] direction ($\theta_p-\theta_p^0=\frac{\pi}{2}$).

Meanwhile, Eq. (6) accounts for the diffusion of Ti species, in which $M_{ij}(x_{\mathrm{Ti}})$ is the overall chemical mobility:

$$M_{ij}(x_{\mathrm{Ti}})=\frac{x_{\mathrm{Ti}}(1-x_{\mathrm{Ti}})}{RT}\left((1-x_{\mathrm{Ti}})\,D_{ij,\mathrm{Ti}}^*+x_{\mathrm{Ti}}\,D_{ij,\mathrm{O}}^*\right) \quad (8)$$

Where $D_{ij,Ti}^*$ and $D_{ij,O}^*$ are the self-diffusivities of Ti and oxygen, respectively, both of which are anisotropic with different values in the (001) plane and along the [001] direction of the bulk rutile. It should be noted that due to the sublattice model and the associated physical constraints, $x_{\mathrm{Ti}}$ is confined to a very narrow composition range of $\frac{1}{3}<x_{Ti}<\frac{2}{5}$, where the upper bound ($\frac{2}{5}$) corresponds to $Ti_2O_3$ and the lower bound ($\frac{1}{3}$) corresponds to $TiO_2$. Directly solving Eq. (6) would easily cause $x_{Ti}$ to exceed this range and lead to numerical convergence issues. To avoid this issue, instead of directly evolving $x_{\mathrm{Ti}}$, we define and evolve an intermediate variable $Y = \ln\frac{6-\frac{2}{x_{Ti}}}{\frac{2}{x_{Ti}}-5}$, which can vary between -∞ and +∞ with $x_{\mathrm{Ti}}$ still satisfying $\frac{1}{3} < x_{Ti} < \frac{2}{5}$. This numerical treatment is detailed in Supplementary Materials B.

These governing equations are solved using an in-house finite-difference program with MPI parallelization. By numerically solving these equations, we can obtain the $TiO_2$ nanocrystal morphologies (from the distribution of $\{\xi_i\}$) and Ti composition distributions at different stages of the nanocarving process.

**Acknowledgement**

The authors acknowledge the start-up fund from College of Engineering, The Ohio State University. Computation is supported by the Ohio Supercomputer Center.

**Funding**

This study was funded by the start-up fund from College of Engineering, The Ohio State University.

**Author contributions**

Yanzhou Ji conceived and supervised the project, developed the overall research framework, contributed to model formulation and data interpretation, and acquired funding and computation resources. Alireza Seifi implemented the computational model, performed simulations, analyzed the results, and prepared figures. Sheikh Akbar contributed to research conceptualization and validation.

Alireza Seifi and Yanzhou Ji wrote the original manuscript. All authors reviewed, edited and approved the final manuscript.

**Conflict of interest statement**

On behalf of all authors, the corresponding author states that there is no conflict of interest.

**Data availability statement**

All data generated or analyzed during this study are included in this published article (and its supplementary information files). The raw data underlying the figures and tables are available from the corresponding author upon reasonable request.

**Supplementary Materials**

# Phase-field modeling of $TiO_2$ nanocarving via reaction with hydrogen-bearing gas

Alireza Seifi, Sheikh Akbar, Yanzhou Ji*

*Department of Materials Science and Engineering, The Ohio State University, Columbus, OH 43210, USA*

## Table of Contents

## A. Thermodynamics of rutile $TiO_2$

The rutile $TiO_2$ phase is described by the sublattice model $(Ti^{3+}, Ti^{4+})_1(O^{2-}, Va)_2$ to account for its nonstoichiometry [30] as the result of the formation of $Ti^{3+}$ and oxygen vacancy during $TiO_2$ reduction reactions. By using this model, it is feasible to capture the thermodynamics of the components, species, endmembers, and the phase itself, as detailed below.

### A.1 Gibbs free energy expressions

Based on, the Gibbs free energy per mole of formula unit, $G_M$, of $TiO_2$ is formulated using the compound energy formalism (CEF):

$$G_M(y^I_{\mathrm{Ti}^{3+}}, y^I_{\mathrm{Ti}^{4+}}, y^{II}_{\mathrm{O}^{2-}}, y^{II}_{\mathrm{Va}}, T) = \sum_{i,j} G^\circ_{i:j}\, y^I_i\, y^{II}_j + RT\,[\sum_i y^I_i \ln y^I_i + 2\sum_j y^{II}_j \ln y^{II}_j] \quad \text{(S1)}$$

Where $G^\circ_{i:j}$ are standard Gibbs free energies of the endmembers, taken from existing database [30].

In the expanded form:

$$\begin{aligned} G_M = \;& G^\circ_{\mathrm{Ti}^{3+}:\mathrm{O}^{2-}} y^I_{\mathrm{Ti}^{3+}} y^{II}_{\mathrm{O}^{2-}} + G^\circ_{\mathrm{Ti}^{4+}:\mathrm{O}^{2-}} y^I_{\mathrm{Ti}^{4+}} y^{II}_{\mathrm{O}^{2-}} \\ & + G^\circ_{\mathrm{Ti}^{3+}:\mathrm{Va}} y^I_{\mathrm{Ti}^{3+}} y^{II}_{\mathrm{Va}} + G^\circ_{\mathrm{Ti}^{4+}:\mathrm{Va}} y^I_{\mathrm{Ti}^{4+}} y^{II}_{\mathrm{Va}} \\ & + RT\,(y^I_{\mathrm{Ti}^{3+}} \ln y^I_{\mathrm{Ti}^{3+}} + y^I_{\mathrm{Ti}^{4+}} \ln y^I_{\mathrm{Ti}^{4+}}) \\ & + 2RT\,(y^{II}_{\mathrm{O}^{2-}} \ln y^{II}_{\mathrm{O}^{2-}} + y^{II}_{\mathrm{Va}} \ln y^{II}_{\mathrm{Va}}) \end{aligned} \quad \text{(S2)}$$

The Gibbs free energy per mole atom of $TiO_2$, $G_m$, can be converted from $G_M$ by dividing $G_M$ with the total number $N$ of atoms in a formula unit:

$$N = y^I_{\mathrm{Ti}^{3+}} + y^I_{\mathrm{Ti}^{4+}} + 2y^{II}_{\mathrm{O}^{2-}} = 1 + 2y^{II}_{\mathrm{O}^{2-}} \quad \text{(S3)}$$

Since vacancy does not count as atoms, its site fraction is not considered in $N$. Therefore:

$$G_m = \frac{G_M}{N} = \frac{G_M}{1 + 2y^{II}_{O^{2-}}} \quad \text{(S4)}$$

### A.2 Chemical Potentials in $TiO_2$

#### *A.2.1 Chemical potentials of endmembers*

According to the sublattice model of rutile $TiO_2$, there are four endmembers: $(Ti^{3+})_1(O^{2-})_2$, $(Ti^{4+})_1(O^{2-})_2$, $(Ti^{3+})_1(Va)_2$, and $(Ti^{4+})_1(Va)_2$, which can also be represented by $Ti^{3+}:O^{2-}$, $Ti^{4+}:O^{2-}$, $Ti^{3+}:Va$, and $Ti^{4+}:Va$, respectively. Their chemical potentials can be calculated by

$$\hat{\mu}_X = G_M + \sum_{i \in X^{(j)}, j=1}^{s} \left(\frac{\partial G_M}{\partial y_i^{(j)}}\right)_{T,p} - \sum_{j=1}^{s} \sum_{i=1}^{m_j} y_i^{(j)} \left(\frac{\partial G_M}{\partial y_i^{(j)}}\right)_{T,p} \quad \text{(S5)}$$

If we define the shorthand derivatives:

$$\Phi^{I}_{Ti^{3+}} \equiv \frac{\partial G_M}{\partial y^{I}_{Ti^{3+}}}, \Phi^{I}_{Ti^{4+}} \equiv \frac{\partial G_M}{\partial y^{I}_{Ti^{4+}}}, \Phi^{II}_{O^{2-}} \equiv \frac{\partial G_M}{\partial y^{II}_{O^{2-}}}, \Phi^{II}_{Va} \equiv \frac{\partial G_M}{\partial y^{II}_{Va}} \quad \text{(S6)}$$

and also define the composition-weighted average derivative term:

$$\langle \Phi \rangle = y^{I}_{Ti^{3+}}\Phi^{I}_{Ti^{3+}} + y^{I}_{Ti^{4+}}\Phi^{I}_{Ti^{4+}} + y^{II}_{O^{2-}}\Phi^{II}_{O^{2-}} + y^{II}_{Va}\Phi^{II}_{Va} \quad \text{(S7)}$$

We can then write down the chemical potentials of these endmembers as below:

1) Endmember $X = (Ti^{3+})_1(O^{2-})_2$

$$\langle \hat{\mu}_{Ti^{3+}:O^{2-}} = G_M + \Phi^{I}_{Ti^{3+}} + \Phi^{II}_{O^{2-}} - \langle \Phi \rangle \quad \text{(S8)}$$

2) Endmember $X = (Ti^{4+})_I : (O^{2-})_{II}$

| | |
|---|---|
| $\hat{\mu}_{\mathrm{Ti^{4+}:O^{2-}}} = G_M + \Phi^{I}_{\mathrm{Ti^{4+}}} + \Phi^{II}_{\mathrm{O^{2-}}} - \langle\Phi\rangle$ | (S9) |

3) Endmember $X = (\mathrm{Ti}^{3+})_I : (\mathrm{Va})_{II}$

| | |
|---|---|
| $\hat{\mu}_{\mathrm{Ti^{3+}:Va}} = G_M + \Phi^{I}_{\mathrm{Ti^{3+}}} + \Phi^{II}_{\mathrm{Va}} - \langle\Phi\rangle$ | (S10) |

4) Endmember $X = (\mathrm{Ti}^{4+})_I : (\mathrm{Va})_{II}$

| | |
|---|---|
| $\hat{\mu}_{\mathrm{Ti^{4+}:Va}} = G_M + \Phi^{I}_{\mathrm{Ti^{4+}}} + \Phi^{II}_{\mathrm{Va}} - \langle\Phi\rangle$ | (S11) |

### *A.2.2 Chemical potentials of sublattice species*

According to the sublattice model of rutile $TiO_2$, there are four sublattice species, i.e,, $(Ti^{3+})^I$, $(Ti^{4+})^I$, $(O^{2-})^{II}$, and $Va^{II}$. Their chemical potentials can be calculated by

| | |
|---|---|
| $\mu_i^{(j)} = \frac{G_M}{\sum_{j=1}^{s} k_j} + \frac{1}{k_j}[(\frac{\partial G_M}{\partial y_i^j})_{T,p,\ y_k^{(l)} \neq y_i^{(j)}} - \sum_{l=1}^{m_j} y_l^{(j)} (\frac{\partial G_M}{\partial y_l^{(j)}})_{T,p,\ y_k^{(m)} \neq y_l^{(j)}}]$ | (S12) |

We first calculate the derivatives:

For $(Ti^{3+})^I$:

| | |
|---|---|
| $\frac{\partial G_M(y^I_{\mathrm{Ti^{3+}}}, y^I_{\mathrm{Ti^{4+}}}, y^{II}_{\mathrm{O^{2-}}}, y^{II}_{\mathrm{Va}})}{\partial y^I_{\mathrm{Ti^{3+}}}} = G^{\circ}_{\mathrm{Ti^{3+}:O^{2-}}}\, y^{II}_{\mathrm{O^{2-}}} + G^{\circ}_{\mathrm{Ti^{3+}:Va}}\, y^{II}_{\mathrm{Va}} + RT(1 + \ln y^I_{\mathrm{Ti^{3+}}})$ | (S13) |

For $(Ti^{4+})^I$:

| | |
|---|---|
| $\frac{\partial G_M(y^I_{\mathrm{Ti^{3+}}}, y^I_{\mathrm{Ti^{4+}}}, y^{II}_{\mathrm{O^{2-}}}, y^{II}_{\mathrm{Va}})}{\partial y^I_{\mathrm{Ti^{4+}}}} = G^{\circ}_{\mathrm{Ti^{4+}:O^{2-}}}\, y^{II}_{\mathrm{O^{2-}}} + G^{\circ}_{\mathrm{Ti^{4+}:Va}}\, y^{II}_{\mathrm{Va}} + RT(1 + \ln y^I_{\mathrm{Ti^{4+}}})$ | (S14) |

For $(O^{2-})^{II}$:

| | |
|---|---|
| $$\frac{\partial G_M(y^I_{\mathrm{Ti}^{3+}}, y^I_{\mathrm{Ti}^{4+}}, y^{II}_{\mathrm{O}^{2-}}, y^{II}_{\mathrm{Va}})}{y^{II}_{\mathrm{O}^{2-}}} = G^{\circ}_{\mathrm{Ti}^{3+}:\mathrm{O}^{2-}}\, y^I_{\mathrm{Ti}^{3+}} + G^{\circ}_{\mathrm{Ti}^{4+}:\mathrm{O}^{2-}}\, y^I_{\mathrm{Ti}^{4+}} + 2RT(1 + \ln y^{II}_{\mathrm{O}^{2-}})$$ | (S15) |

For $Va^{II}$:

| | |
|---|---|
| $$\frac{\partial G_M(y^I_{\mathrm{Ti}^{3+}}, y^I_{\mathrm{Ti}^{4+}}, y^{II}_{\mathrm{O}^{2-}}, y^{II}_{\mathrm{Va}})}{y^{II}_{\mathrm{Va}}} = G^{\circ}_{\mathrm{Ti}^{3+}:\mathrm{Va}}\, y^I_{\mathrm{Ti}^{3+}} + G^{\circ}_{\mathrm{Ti}^{4+}:\mathrm{Va}}\, y^I_{\mathrm{Ti}^{4+}} + 2RT(1 + \ln y^{II}_{\mathrm{Va}})$$ | (S16) |

We can then find their chemical potentials:

Chemical potential of $Ti^{3+}$ on sublattice I:

| | |
|---|---|
| $$\mu^I_{\mathrm{Ti}^{3+}} = \frac{1}{3}[G^{\circ}_{\mathrm{Ti}^{3+}:\mathrm{O}^{2-}}\, y^{II}_{\mathrm{O}^{2-}}\,(3 - 2y^I_{\mathrm{Ti}^{3+}}) + G^{\circ}_{\mathrm{Ti}^{3+}:\mathrm{Va}}\, y^{II}_{\mathrm{Va}}\,(3 - 2y^I_{\mathrm{Ti}^{3+}}) -2\, y^I_{\mathrm{Ti}^{4+}}(G^{\circ}_{\mathrm{Ti}^{4+}:\mathrm{O}^{2-}}\, y^{II}_{\mathrm{O}^{2-}} + G^{\circ}_{\mathrm{Ti}^{4+}:\mathrm{Va}}\, y^{II}_{\mathrm{Va}})] +RT[2\, y^{II}_{\mathrm{O}^{2-}} \ln y^{II}_{\mathrm{O}^{2-}} + (3 - 2y^I_{\mathrm{Ti}^{3+}}) \ln y^I_{\mathrm{Ti}^{3+}} -2\, y^I_{\mathrm{Ti}^{4+}} \ln y^I_{\mathrm{Ti}^{4+}} + 2\, y^{II}_{\mathrm{Va}} \ln y^{II}_{Va} - 3 - y^I_{\mathrm{Ti}^{3+}} + y^I_{\mathrm{Ti}^{4+}}]$$ | (S17) |

Chemical potential of $Ti^{4+}$ on sublattice I:

| | |
|---|---|
| $$\mu^I_{\mathrm{Ti}^{4+}} = \frac{1}{3}[G^{\circ}_{\mathrm{Ti}^{4+}:\mathrm{O}^{2-}}\, y^{II}_{\mathrm{O}^{2-}}\left(3 - 2y^I_{\mathrm{Ti}^{4+}}\right) + G^{\circ}_{\mathrm{Ti}^{4+}:\mathrm{Va}}\, y^{II}_{\mathrm{Va}}\left(3 - 2y^I_{\mathrm{Ti}^{4+}}\right) - 2\, y^I_{\mathrm{Ti}^{3+}}(G^{\circ}_{\mathrm{Ti}^{3+}:\mathrm{O}^{2-}}\, y^{II}_{\mathrm{O}^{2-}} + G^{\circ}_{\mathrm{Ti}^{3+}:\mathrm{Va}}\, y^{II}_{\mathrm{Va}}) +RT[2\, y^{II}_{\mathrm{O}^{2-}} \ln y^{II}_{\mathrm{O}^{2-}} + \left(3 - 2y^I_{\mathrm{Ti}^{4+}}\right) \ln y^I_{\mathrm{Ti}^{4+}} - 2\, y^I_{\mathrm{Ti}^{3+}} \ln y^I_{\mathrm{Ti}^{3+}} + 2\, y^{II}_{\mathrm{Va}} \ln y^{II}_{\mathrm{Va}} - 3 + y^I_{\mathrm{Ti}^{3+}} - y^I_{\mathrm{Ti}^{4+}}]]$$ | (S18) |

Chemical potential of $O^{2-}$ on sublattice II:

| | |
|---|---|
| $$\mu^{II}_{\mathrm{O}^{2-}} = \frac{1}{6}[-((-3 + y^{II}_{\mathrm{O}^{2-}})(G^{\circ}_{\mathrm{Ti}^{3+}:\mathrm{O}^{2-}}\, y^I_{\mathrm{Ti}^{3+}} + G^{\circ}_{\mathrm{Ti}^{4+}:\mathrm{O}^{2-}}\, y^I_{\mathrm{Ti}^{4+}})) - (G^{\circ}_{\mathrm{Ti}^{3+}:\mathrm{Va}}\, y^I_{\mathrm{Ti}^{3+}} + G^{\circ}_{\mathrm{Ti}^{4+}:\mathrm{Va}}\, y^I_{\mathrm{Ti}^{4+}})\, y^{II}_{\mathrm{Va}} -6RT\left(-1 + y^{II}_{\mathrm{O}^{2-}} + y^{II}_{\mathrm{Va}}\right) + 2RT(-\left(\left(-3 + y^{II}_{\mathrm{O}^{2-}}\right) \ln y^{II}_{\mathrm{O}^{2-}}\right) +y^I_{\mathrm{Ti}^{3+}} \ln y^I_{\mathrm{Ti}^{3+}} + y^I_{\mathrm{Ti}^{4+}} \ln y^I_{\mathrm{Ti}^{4+}} - y^{II}_{\mathrm{Va}} \ln y^{II}_{\mathrm{Va}})]$$ | (S19) |

Chemical potential of Va on sublattice II:

| | |
|---|---|
| $\mu_{\mathrm{Va}}^{II} = \frac{1}{6}[3\, G^{\circ}_{\mathrm{Ti^{4+}:Va}}\, y^{I}_{\mathrm{Ti^{4+}}} - y^{II}_{\mathrm{O^{2-}}}(G^{\circ}_{\mathrm{Ti^{3+}:O^{2-}}}\, y^{I}_{\mathrm{Ti^{3+}}} + G^{\circ}_{\mathrm{Ti^{4+}:O^{2-}}}\, y^{I}_{\mathrm{Ti^{4+}}}) - G^{\circ}_{\mathrm{Ti^{3+}:Va}}\, y^{I}_{\mathrm{Ti^{3+}}}\,(-3 + y^{II}_{\mathrm{Va}}) - G^{\circ}_{\mathrm{Ti^{4+}:Va}}\, y^{I}_{\mathrm{Ti^{4+}}}\, y^{II}_{\mathrm{Va}} - 6RT\left(-1 + y^{II}_{\mathrm{O^{2-}}} + y^{II}_{\mathrm{Va}}\right) + 2RT(-y^{II}_{\mathrm{O^{2-}}} \ln y^{II}_{\mathrm{O^{2-}}} + y^{I}_{\mathrm{Ti^{3+}}} \ln y^{I}_{\mathrm{Ti^{3+}}} + y^{I}_{\mathrm{Ti^{4+}}} \ln y^{I}_{\mathrm{Ti^{4+}}} - (-3 + y^{II}_{\mathrm{Va}}) \ln y^{II}_{\mathrm{Va}})]$ | (S20) |

### *A.2.3 Chemical potentials of components and species*

The rutile $TiO_2$ phase has two components, Ti and O, and four species, $Ti^{3+}$, $Ti^{4+}$, $O^{2-}$, and Va. Note that the species here are defined with respect to the whole phase, which is different from the "sublattice species" in the previous section defined on given sublattices. However, since the Gibbs free energy of rutile is not directly formulated as a function of their compositions, their chemical potentials are not as straightforward. Here we discuss their expressions in detail.

Based on stoichiometric thermodynamics, we have:

| | |
|---|---|
| $\mu_{\mathrm{Ti}} + 2\mu_{\mathrm{O}} + \mu_e = \hat{\mu}_{\mathrm{Ti^{3+}:O^{2-}}} = \mu_{(\mathrm{Ti^{3+}})_1(\mathrm{O^{2-}})_2}$ | (S21a) |
| $\mu_{\mathrm{Ti}} + 2\mu_{\mathrm{O}} = \hat{\mu}_{\mathrm{Ti^{4+}:O^{2-}}} = \mu_{(\mathrm{Ti^{4+}})_1(\mathrm{O^{2-}})_2}$ | (S21b) |

Therefore, we can obtain the chemical potential of electrons:

| | |
|---|---|
| $\mu_e = \mu_{(\mathrm{Ti^{3+}})_1(\mathrm{O^{2-}})_2} - \mu_{(\mathrm{Ti^{4+}})_1(\mathrm{O^{2-}})_2}$ | (S22) |

Also, according to the relation between endmember and sublattice chemical potentials

| | |
|---|---|
| $\mu_{(\mathrm{Ti^{3+}})_1(\mathrm{O^{2-}})_2} = \mu^{I}_{\mathrm{Ti^{3+}}} + 2\mu^{II}_{\mathrm{O^{2-}}}$ | (S23a) |
| $\mu_{(\mathrm{Ti^{4+}})_1(\mathrm{O^{2-}})_2} = \mu^{I}_{\mathrm{Ti^{4+}}} + 2\mu^{II}_{\mathrm{O^{2-}}}$ | (S23b) |

We have

| | |
|---|---|
| $\mu_e = \mu^{I}_{\mathrm{Ti^{3+}}} - \mu^{I}_{\mathrm{Ti^{4+}}}$ | (S24) |

Therefore, using the $\mu^{I}_{\mathrm{Ti}^{3+}}$ and $\mu^{I}_{\mathrm{Ti}^{4+}}$ expressions derived from previous sections, we obtain

$$\mu_e = y^{II}_{\mathrm{O}^{2-}}(G^{\circ}_{\mathrm{Ti}^{3+}:\mathrm{O}^{2-}} - G^{\circ}_{\mathrm{Ti}^{4+}:\mathrm{O}^{2-}}) + y^{II}_{\mathrm{Va}}(G^{\circ}_{\mathrm{Ti}^{3+}:\mathrm{Va}} - G^{\circ}_{\mathrm{Ti}^{4+}:\mathrm{Va}}) + RT\ln(\frac{y^{I}_{\mathrm{Ti}^{3+}}}{y^{I}_{\mathrm{Ti}^{4+}}}) \quad \text{(S25)}$$

Now, we can derive the chemical potential of Ti:

$$\mu_{(\mathrm{Ti}^{3+})_1(\mathrm{Va})_2} = \hat{\mu}_{\mathrm{Ti}^{3+}:\mathrm{Va}} = \mu_{\mathrm{Ti}} - 3\mu_e = \mu^{I}_{\mathrm{Ti}^{3+}} + 2\mu^{II}_{\mathrm{Va}}$$
$$\Rightarrow \ \mu_{Ti} = \mu^{I}_{\mathrm{Ti}^{3+}} + 2\mu^{II}_{\mathrm{Va}} + 3\mu_e \quad \text{(S26)}$$

using the $\mu^{I}_{\mathrm{Ti}^{3+}}$, $\mu^{II}_{\mathrm{Va}}$ and $\mu_e$ expressions derived above, we obtain

$$\begin{aligned}\mu_{Ti} = {} & -3\, G^{\circ}_{\mathrm{Ti}^{4+}:\mathrm{O}^{2-}}\, y^{II}_{\mathrm{O}^{2-}} - G^{\circ}_{\mathrm{Ti}^{3+}:\mathrm{O}^{2-}}\, y^{II}_{\mathrm{O}^{2-}}\left(-4 + y^{I}_{\mathrm{Ti}^{3+}}\right) + G^{\circ}_{\mathrm{Ti}^{3+}:\mathrm{Va}}\, y^{I}_{\mathrm{Ti}^{3+}} + \\ & G^{\circ}_{\mathrm{Ti}^{4+}:\mathrm{Va}}\, y^{I}_{\mathrm{Ti}^{4+}} - G^{\circ}_{\mathrm{Ti}^{4+}:\mathrm{O}^{2-}}\, y^{II}_{\mathrm{O}^{2-}}\, y^{I}_{\mathrm{Ti}^{4+}} - \\ & (G^{\circ}_{\mathrm{Ti}^{3+}:\mathrm{Va}}(-4 + y^{I}_{\mathrm{Ti}^{3+}}) + G^{\circ}_{\mathrm{Ti}^{4+}:\mathrm{Va}}(3 + y^{I}_{\mathrm{Ti}^{4+}}))\, y^{II}_{\mathrm{Va}} \\ & - RT(-3 + 2y^{II}_{\mathrm{O}^{2-}} + y^{I}_{\mathrm{Ti}^{3+}} + y^{I}_{\mathrm{Ti}^{4+}} + 2y^{II}_{\mathrm{Va}}) \\ & + RT(4\ln y^{I}_{\mathrm{Ti}^{3+}} - 3\ln y^{I}_{\mathrm{Ti}^{4+}} + 2\ln y^{II}_{\mathrm{Va}})\end{aligned} \quad \text{(S27)}$$

Also for oxygen we can write:

$$\mu_{\mathrm{Ti}} + 2\mu_{\mathrm{O}} = \mu_{(\mathrm{Ti}^{4+})_1(\mathrm{O}^{2-})_2} \quad \text{(S28)}$$

From this, we can obtain the chemical potential of O:

$$\mu_{\mathrm{O}} = \frac{1}{2}\left(\mu^{I}_{\mathrm{Ti}^{4+}} + 2\mu^{II}_{\mathrm{O}^{2-}} - \mu_{\mathrm{Ti}}\right) \quad \text{(S29)}$$

Which has the detailed expression of

| | |
|---|---|
| $$\mu_O = \frac{1}{2}[4\, G^{\circ}_{Ti^{4+}:O^{2-}}\, y^{II}_{O^{2-}} + G^{\circ}_{Ti^{3+}:O^{2-}}\,(-4\, y^{II}_{O^{2-}} + y^{I}_{Ti^{3+}}) + G^{\circ}_{Ti^{4+}:O^{2-}}\, y^{I}_{Ti^{4+}} - G^{\circ}_{Ti^{4+}:Va}\, y^{I}_{Ti^{4+}} + 4\, G^{\circ}_{Ti^{4+}:Va}\, y^{II}_{Va} - G^{\circ}_{Ti^{3+}:Va}\,(y^{I}_{Ti^{3+}} + 4\, y^{II}_{Va}) + +2RT(ln\, y^{II}_{O^{2-}} - 2ln\, y^{I}_{Ti^{3+}} + 2ln\, y^{I}_{Ti^{4+}} - ln\, y^{II}_{Va})]$$ | (S30) |

We can then obtain the species chemical potential expressions as

| | |
|---|---|
| $$\mu_{\mathrm{Ti}^{3+}} = \mu_{\mathrm{Ti}} - 3\mu_e = G^{\circ}_{\mathrm{Ti}^{4+}:\mathrm{Va}}\, y^{I}_{\mathrm{Ti}^{4+}} + y^{II}_{\mathrm{O}^{2-}}(G^{\circ}_{\mathrm{Ti}^{3+}:\mathrm{O}^{2-}} - G^{\circ}_{\mathrm{Ti}^{3+}:\mathrm{O}^{2-}}\, y^{I}_{\mathrm{Ti}^{3+}} - G^{\circ}_{\mathrm{Ti}^{4+}:\mathrm{O}^{2-}}\, y^{I}_{\mathrm{Ti}^{4+}}) - (G^{\circ}_{\mathrm{Ti}^{3+}:\mathrm{Va}}\, y^{I}_{\mathrm{Ti}^{3+}} + G^{\circ}_{\mathrm{Ti}^{4+}:\mathrm{Va}}\, y^{I}_{\mathrm{Ti}^{4+}})\, y^{II}_{\mathrm{Va}} + G^{\circ}_{\mathrm{Ti}^{3+}:\mathrm{Va}}(y^{I}_{\mathrm{Ti}^{3+}} + y^{II}_{\mathrm{Va}}) - RT(-3 + 2y^{II}_{\mathrm{O}^{2-}} + y^{I}_{\mathrm{Ti}^{3+}} + y^{I}_{\mathrm{Ti}^{4+}} + 2y^{II}_{\mathrm{Va}}) + RT(\ln y^{I}_{\mathrm{Ti}^{3+}} + 2\ln y^{II}_{\mathrm{Va}}).$$ | (S31) |
| $$\mu_{\mathrm{Ti}^{4+}} = \mu_{\mathrm{Ti}} - 4\mu_e = G^{\circ}_{\mathrm{Ti}^{3+}:\mathrm{Va}}\, y^{I}_{\mathrm{Ti}^{3+}} + y^{II}_{\mathrm{O}^{2-}}(G^{\circ}_{\mathrm{Ti}^{4+}:\mathrm{O}^{2-}} - G^{\circ}_{\mathrm{Ti}^{3+}:\mathrm{O}^{2-}}\, y^{I}_{\mathrm{Ti}^{3+}} - G^{\circ}_{\mathrm{Ti}^{4+}:\mathrm{O}^{2-}}\, y^{I}_{\mathrm{Ti}^{4+}}) - (G^{\circ}_{\mathrm{Ti}^{3+}:\mathrm{Va}}\, y^{I}_{\mathrm{Ti}^{3+}} + G^{\circ}_{\mathrm{Ti}^{4+}:\mathrm{Va}}\, y^{I}_{\mathrm{Ti}^{4+}})\, y^{II}_{\mathrm{Va}} + G^{\circ}_{\mathrm{Ti}^{4+}:\mathrm{Va}}(y^{I}_{\mathrm{Ti}^{4+}} + y^{II}_{\mathrm{Va}}) - RT(-3 + 2y^{II}_{\mathrm{O}^{2-}} + y^{I}_{\mathrm{Ti}^{3+}} + y^{I}_{\mathrm{Ti}^{4+}} + 2y^{II}_{\mathrm{Va}}) + RT(\ln y^{I}_{\mathrm{Ti}^{4+}} + 2\ln y^{II}_{\mathrm{Va}})$$ | (S32) |
| $$\mu_{\mathrm{O}^{2-}} = \mu_{\mathrm{O}} + 2\mu_e = \frac{1}{6}[3\, G^{\circ}_{\mathrm{Ti}^{3+}:\mathrm{O}^{2-}}\, y^{I}_{\mathrm{Ti}^{3+}} - G^{\circ}_{\mathrm{Ti}^{3+}:\mathrm{O}^{2-}}\, y^{II}_{\mathrm{O}^{2-}}\,(12 + y^{I}_{\mathrm{Ti}^{3+}}) + G^{\circ}_{\mathrm{Ti}^{4+}:\mathrm{O}^{2-}}(-\, y^{II}_{\mathrm{O}^{2-}}\,(-12 + y^{I}_{\mathrm{Ti}^{4+}}) + 3\, y^{I}_{\mathrm{Ti}^{4+}}) -(G^{\circ}_{\mathrm{Ti}^{3+}:\mathrm{Va}}\,(12 + y^{I}_{\mathrm{Ti}^{3+}}) + G^{\circ}_{\mathrm{Ti}^{4+}:\mathrm{Va}}\,(-12 + y^{I}_{\mathrm{Ti}^{4+}}))\, y^{II}_{\mathrm{Va}} - 6RT\,(-1 + y^{II}_{\mathrm{O}^{2-}} + y^{II}_{\mathrm{Va}}) + 2RT(-((-3 + y^{II}_{\mathrm{O}^{2-}})\ln y^{II}_{\mathrm{O}^{2-}}) + (-6 + y^{I}_{\mathrm{Ti}^{3+}})\ln y^{I}_{\mathrm{Ti}^{3+}} +(6 + y^{I}_{\mathrm{Ti}^{4+}})\ln y^{I}_{\mathrm{Ti}^{4+}} - y^{II}_{\mathrm{Va}}\ln y^{II}_{\mathrm{Va}})]$$ | (S33) |
| $$\mu_{\mathrm{Va}} = 0$$ | (S34) |

### A.3 Conversion of Sublattice Site Fractions to Molar Fraction of Ti

The site fractions of sublattice species in the rutile phase can be converted into Ti composition. Here the Ti composition contains contributions from both $Ti^{3+}$ and $Ti^{4+}$. Following the procedures in [22, 32], we list all the constraints for site fractions, including site conservations, mass balance, and charge neutrality:

| | |
|---|---|
| $y^{I}_{\mathrm{Ti}^{3+}} + y^{I}_{\mathrm{Ti}^{4+}} = 1$ | (S35a) |
| $y^{II}_{\mathrm{O}^{2-}} + y^{II}_{\mathrm{Va}} = 1$ | (S35b) |
| $3\, y^{I}_{\mathrm{Ti}^{3+}} + 4\, y^{I}_{\mathrm{Ti}^{4+}} = 4\, y^{II}_{\mathrm{O}^{2-}}$ | (S35c) |
| $x_{\mathrm{Ti}} = \dfrac{1}{1 + 2\, y^{II}_{\mathrm{O}^{2-}}}$ | (S35d) |

From these equations, we can solve for the site fractions as a function of $x_{\mathrm{Ti}}$:

| | |
|---|---|
| $y^{I}_{\mathrm{Ti}^{3+}} = 6 - \dfrac{2}{x_{\mathrm{Ti}}}$ | (S36a) |
| $y^{I}_{\mathrm{Ti}^{4+}} = \dfrac{2}{x_{\mathrm{Ti}}} - 5$ | (S36b) |
| $y^{II}_{\mathrm{O}^{2-}} = \dfrac{1}{2}\,(-1 + \dfrac{1}{x_{\mathrm{Ti}}})$ | (S36c) |
| $y^{II}_{\mathrm{Va}} = \dfrac{3}{2} - \dfrac{1}{2x_{\mathrm{Ti}}}$ | (S36d) |

With these relations, the Gibbs free energy expressions and all the chemical potentials mentioned above can be converted into functions of Ti composition. The resulting thermodynamic functions at T=1000 K are shown in Figure S1.

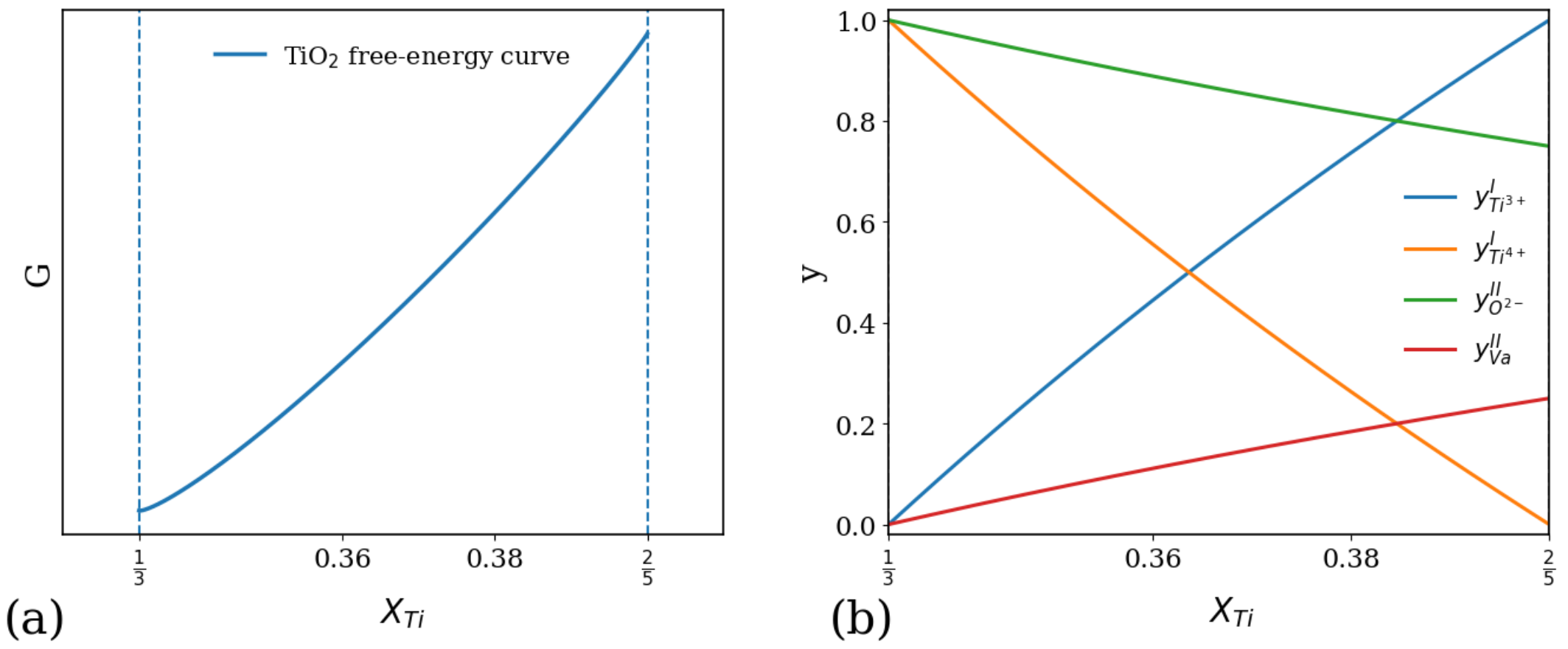

Figure S1: (a) $TiO_2$ molar Gibbs free energy as a function of Ti composition at T = 1000 K; (b) Sublattice site fractions as functions of Ti composition.

### A.4 $\Delta G_m^r$ expressions

Now we look into the expressions for the Gibbs free energy change of the $TiO_2$ reduction reaction, $\Delta G_m^r$. This reduction can be written in different forms, and their corresponding $\Delta G_m^r$ expressions are all consistent. The reaction happening in the system is:

$$2Ti^{4+} + H_2 + O^{2-} = 2Ti^{3+} + H_2O \quad \text{(S37)}$$

We start from:

$$\Delta G_m^r = 2\mu_{Ti^{3+}} - 2\mu_{Ti^{4+}} - \mu_{O^{2-}} + \mu_{H_2O} - \mu_{H_2} \quad \text{(S38)}$$

From our previous sections:

$$\mu_{\mathrm{Ti}^{3+}} = \mu_{\mathrm{Ti}} - 3\mu_e, \mu_{\mathrm{Ti}^{4+}} = \mu_{\mathrm{Ti}} - 4\mu_e, \mu_{\mathrm{O}^{2-}} = \mu_{\mathrm{O}} + 2\mu_e, \mu_{\mathrm{Va}} = 0. \quad \text{(S39)}$$

Now inserting in $\Delta G_m^r$:

$$\begin{aligned} \Delta G_m^r &= 2(\mu_{\mathrm{Ti}} - 3\mu_e) - 2(\mu_{\mathrm{Ti}} - 4\mu_e) - (\mu_{\mathrm{O}} + 2\mu_e) + \mu_{H_2O} - \mu_{H_2} \\ &= 2\mu_{\mathrm{Ti}} - 6\mu_e - 2\mu_{\mathrm{Ti}} + 8\mu_e - \mu_{\mathrm{O}} - 2\mu_e + \mu_{H_2O} - \mu_{H_2} \\ &= -\mu_{\mathrm{O}} + \mu_{H_2O} - \mu_{H_2} \end{aligned} \quad \text{(S40)}$$

Based on the relationships among component chemical potentials, sublattice species chemical potentials, and endmember chemical potentials discussed above, we can also have

$$\Delta G_m^r = \frac{1}{2}\mu_{(\mathrm{Ti}^{3+})_1(\mathrm{Va})_2} + \frac{3}{2}\mu_{(\mathrm{Ti}^{3+})_1(\mathrm{O}^{2-})_2} - 2\mu_{(\mathrm{Ti}^{4+})_1(\mathrm{O}^{2-})_2} + \mu_{H_2O} - \mu_{H_2} \tag{S41}$$

And

$$\Delta G_m^r = 2\mu_{\mathrm{Ti}^{3+}}^{I} + \mu_{\mathrm{Va}}^{II} - 2\mu_{\mathrm{Ti}^{4+}}^{I} - \mu_{\mathrm{O}^{2-}}^{II} + \mu_{H_2O} - \mu_{H_2} \tag{S42}$$

These expressions are all consistent. To sum up,

$$\begin{aligned}\Delta G_m^r &= 2\mu_{Ti^{3+}} - 2\mu_{Ti^{4+}} - \mu_{O^{2-}} + \mu_{H_2O} - \mu_{H_2} \\ &= 2\mu_{\mathrm{Ti}^{3+}}^{I} + \mu_{\mathrm{Va}}^{II} - 2\mu_{\mathrm{Ti}^{4+}}^{I} - \mu_{\mathrm{O}^{2-}}^{II} + \mu_{H_2O} - \mu_{H_2} \\ &= -\mu_{\mathrm{O}} + \mu_{H_2O} - \mu_{H_2} \\ &= \frac{1}{2}\mu_{(\mathrm{Ti}^{3+})_1(\mathrm{Va})_2} + \frac{3}{2}\mu_{(\mathrm{Ti}^{3+})_1(\mathrm{O}^{2-})_2} - 2\mu_{(\mathrm{Ti}^{4+})_1(\mathrm{O}^{2-})_2} + \mu_{H_2O} - \mu_{H_2}\end{aligned} \tag{S43}$$

With the variable conversion in Section A.3 above, the $\Delta G_m^r$ can be expressed as a function of $x_{\mathrm{Ti}}$:

$$\begin{aligned}\Delta G_m^r = &\frac{G^{\circ}_{Ti^{3+}:O^{2-}}(4 - 8x_{Ti}) + x_{Ti}\left(12\, G^{\circ}_{Ti^{3+}:Va} + 7\, G^{\circ}_{Ti^{4+}:O^{2-}} - 11\, G^{\circ}_{Ti^{4+}:Va} + RTln\, 16\right)}{2x_{Ti}} \\ &\frac{-4\left(G^{\circ}_{Ti^{3+}:Va} + G^{\circ}_{Ti^{4+}:O^{2-}} - G^{\circ}_{Ti^{4+}:Va}\right) - 2RTx_{Ti}[2\ln(2 - 5x_{Ti}) + \ln(1 - x_{Ti}) - 3\ln(3x_{Ti} - 1)]}{2x_{Ti}} \\ &+\mu^0_{H_2O} - \mu^0_{H_2} + RTln\left(\frac{P_{H_2O}}{P_{H_2}}\right)\end{aligned} \tag{S44}$$

This is the expression we directly implement in phase-field simulations. We can also plot $\Delta G_m^r$ as a function of $x_{\mathrm{Ti}}$ and $\frac{P_{H_2O}}{P_{H_2}}$ ratio:

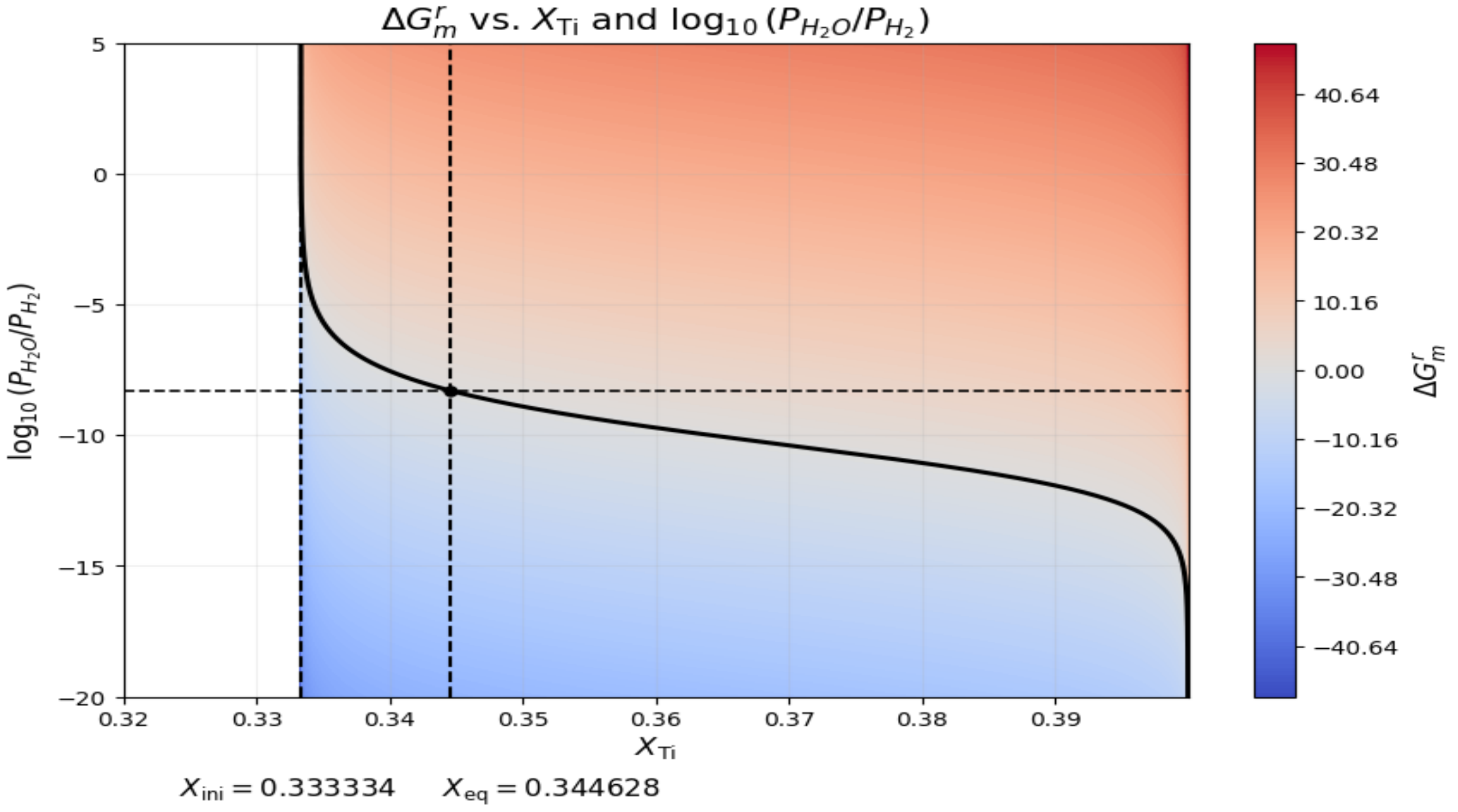

Figure S2: Reaction free energy $\Delta G_m^r$(dimensionless) mapped as a function of titanium composition $x_{Ti}$ and gas-phase ratio $log_{10}(P_{H_2O}/P_{H_2})$ at $T = 1000\ K$. The colormap represents the magnitude and sign of $\Delta G_m^r$, with the zero-contour (black curve) defining thermodynamic equilibrium. Regions with $\Delta G_m^r < 0$(blue) indicate reduction-favorable conditions, while $\Delta G_m^r > 0$(red) correspond to oxidation-favorable conditions. The horizontal dashed line marks the selected gas ratio ($P_{H_2O}/P_{H_2} = 5 \times 10^{-9}$), and its intersection with the equilibrium curve defines the equilibrium composition $x_{eq} = 0.344628$. The initial composition $x_{ini} = 0.333334$ is indicated by the vertical dashed line, highlighting the thermodynamic driving force toward equilibration under the specified conditions.

## B. Intermediate variables for solving the diffusion-reaction equation

As we mentioned in the previous sections, the thermodynamic description of the rutile $TiO_2$ follows a sublattice model $(Ti^{3+}, Ti^{4+})_1(O^{2-}, Va)_2$. Consequently, when we solve the site fractions from $x_{\mathrm{Ti}}$ from equations in section IV., and impose the constraints of $0 \le y_i^{(j)} \le 1$ for all i and j, we have:

$$\frac{1}{3} \le x_{Ti} \le \frac{2}{5} \tag{S45}$$

Which gives us a very narrow boundary for the $x_{Ti}$. To avoid the composition going beyond this boundary, we introduce an intermediate variable Y in which Y can vary between -∞ and +∞ but the $x_{Ti}$ still meets the condition mentioned above. By doing this, it is possible to evolve the intermediate variable Y instead of $x_{Ti}$ in the diffusion-reaction equation. We define the intermediate variable Y as

$$Y = \ln \frac{6 - \frac{2}{x_{Ti}}}{\frac{2}{x_{Ti}} - 5} \tag{S46}$$

So

$$x_{Ti} = \frac{2(1 + e^Y)}{6 + 5e^Y} \tag{S47}$$

Therefore,

$$\frac{\partial x_{Ti}}{\partial t} = \frac{\partial Y}{\partial t} \cdot \frac{\partial x_{Ti}}{\partial Y} = \frac{\partial Y}{\partial t} \cdot \frac{2e^Y}{(6 + 5e^Y)^2} \tag{S48}$$

Plugging it into the diffusion-equation in the main text,

$$\left[1 - \sum_i h(\xi_i)\right] \frac{\partial x_{Ti}}{\partial t} = \nabla \cdot \mathrm{D}(\{\xi_i\}, T) \nabla x_{Ti} + x_{Ti} \sum_i \frac{\partial h}{\partial \xi_i} \frac{\partial \xi_i}{\partial t} \tag{S49}$$

Now the left-hand-side of this equation becomes

$$\begin{aligned} &\left[1 - \sum_i h(\xi_i)\right] \frac{\partial Y}{\partial t} \cdot \frac{2e^Y}{(6 + 5e^Y)^2} + \frac{\partial Y}{\partial t} - \frac{\partial Y}{\partial t} \\ &= \frac{\partial Y}{\partial t} + \left[\left[1 - \sum_i h(\xi_i)\right] \frac{2e^Y}{(6 + 5e^Y)^2} - 1\right] \cdot \frac{\partial Y}{\partial t} \end{aligned} \tag{S50}$$

Moving the term $\left[ [1 - \sum_i h(\xi_i)] \frac{2e^Y}{(6+5e^Y)^2} - 1 \right] . \frac{\partial Y}{\partial t}$ to the right-hand-side, we obtain the governing diffusion-equation that we indeed solve in the phase-field simulation:

$$\frac{\partial Y}{\partial t} = \nabla \cdot \mathrm{D}(\{\xi_i\}, T) \nabla x_{Ti} + x_{Ti} \sum_i \frac{\partial h}{\partial \xi_i} \frac{\partial \xi_i}{\partial t} - \left[ \left[ 1 - \sum_i h(\xi_i) \right] \frac{2e^Y}{(6 + 5e^Y)^2} - 1 \right] . \frac{\partial Y}{\partial t} \quad \text{(S51)}$$